\documentclass[aps,prd,11pt,twoside,tightenlines,nofootinbib,showpacs,preprint,superscriptaddress]{revtex4-1}
\usepackage[colorlinks=true]{hyperref}
\usepackage{xcolor}
\hypersetup{colorlinks=true, citecolor=blue, urlcolor=blue, linkcolor=blue}
\usepackage{amsmath,amssymb}
\usepackage[final]{graphicx}
\usepackage{subcaption}
\usepackage{float}
\usepackage[toc,page]{appendix}

\begin{document}

\title{Study $B^{0}_{(s)}$ decays into $\phi$ and a scalar or vector meson}

\author{Hiwa A. Ahmed}
\affiliation{School of Physics and Electronics, Central South University, Changsha 410083, China}
\affiliation{Physics Department, College of Science, University of Sulaimani, Kurdistan Region 46001, Iraq}

\author{Zhong-Yu Wang}
\affiliation{School of Physics and Electronics, Central South University, Changsha 410083, China}

\author{Zhi-Feng Sun}
\email{sunzf@lzu.edu.cn}
\affiliation{School of Physical Science and Technology, Lanzhou University, Lanzhou 730000, China}

\author{C. W. Xiao}
\email{xiaochw@csu.edu.cn}
\affiliation{School of Physics and Electronics, Central South University, Changsha 410083, China}

\date{\today}

\begin{abstract}

In the present work, we investigate the decays of $B^{0}_{s} \rightarrow \phi\pi^{+}\pi^{-}$ and $B^{0} \rightarrow \phi\pi^{+}\pi^{-}$ with the final state interactions based on the chiral unitary approach. In the final state interactions of the $\pi^+\pi^-$ with its coupled channels, we study the effects of the $\eta\eta$ channel in the two-body interactions for the reproduction of the $f_{0}(980)$ state. Our results for the $\pi^+\pi^-$ invariant mass distributions of the decay $B^{0}_{s} \rightarrow \phi\pi^{+}\pi^{-}$ describe the experimental data up to 1 GeV well, with the resonance contributions from the $f_{0}(980)$ and $\rho$. For the predicted invariant mass distributions of the $B^{0} \rightarrow \phi\pi^{+}\pi^{-}$ decay, we found that the contributions from the $f_{0}(500)$ are significant except for the ones from the $f_{0}(980)$ state. With some experimental branching ratios as input to determine the production vertex factors, we make some predictions for the branching ratios of the other final decay channels, including the vector mesons, in the $B^0_{(s)}$ decays, where some of them are consistent with the experimental ones within the uncertainties.

\end{abstract}

\maketitle

\section{Introduction}

For hunting the $CP$-violating of the new physics effects beyond the Standard Model (SM), the three body $B$ meson decays are caught much attentions both in the theoreties and the experiments. Due to the exceptional progress of the experiments, many non-leptonic three body $B$ meson decays have been measured by the collaboratoins of Belle, BaBar, LHCb and so on \cite{pdg2018}. 
On the other hand, to investigate the three body non-leptonic $B$ meson decays in theories, some theoretical models are proposed, 
such as the $SU(3)$ flavor symmetry framework \cite{Savage:1989ub,Lipkin:1991st,Grossman:2003qp,Gronau:2006qn,Bhattacharya:2014eca,Deshpande:2002be,Xu:2013dta,He:2014xha}, the heavy quark effective theory combined with the chiral perturbation theory and the final state interactions \cite{Deshpande:1995nu,Fajfer:1998yc,Deandrea:2000tf,Deandrea:2000ce,Gardner:2001gc,Cheng:2002qu,Fajfer:2004cx,Bediaga:2008zz,Cheng:2013dua,Li:2014oca,Daub:2015xja,Boito:2017jav}, QCD sum rules under the factorization approach \cite{Leitner:2002xh,Chua:2001vh,Chua:2004mi,Zhang:2013oqa,Mohammadi:2014rpa,Mohammadi:2014eia}, the perturbative QCD approach \cite{Chen:2002th,Chen:2004az,Wang:2014qya,Wang:2015uea,Li:2015tja,Wang:2016rlo,Ma:2016csn,Morales:2016pcq,Li:2016tpn},  the final state interaction formalis \cite{Liang:2014tia,Liang:2015qva,Bayar:2014qha} based on the chiral unitary approach (ChUA) \cite{Oller:1997ti,Oset:1997it,Oller:2000ma,Oller:2000fj,Hyodo:2008xr,Oset:2008qh}, and so on.

Aiming at searching for the effect of physics beyond the SM, the LHCb collaboration have reported the first observation the rare three body decays of $B^{0}_{s} \rightarrow \phi\pi^{+}\pi^{-}$ and $B^{0} \rightarrow \phi\pi^{+}\pi^{-}$ \cite{Aaij:2016qnm} \footnote{Note that, sometimes these decay processes are referred as the $\bar{B}^{0}_{s}$ and $\bar{B}^{0}$ mesons' decays, since that they are not identical in the experimental measurements due to the particle pair productions and the charge symmetry.}, where the $B^{0}_{s} \rightarrow \phi\pi^{+}\pi^{-}$ decay investigated with the requirement of the $\pi^{+}\pi^{-}$ invariant mass in the range $400 \textsl{ MeV} < m(\pi\pi) < 1600 \textsl{ MeV} $, and some resonant contributions from the states of $\rho(770)$, $f_{0}(980)$, $f_{2}(1270)$, and $f_{0}(1500)$ are found in the $m(\pi\pi)$ invariant mass spectrum. The decays of $B^{0}_{s} \rightarrow \phi\pi^{+}\pi^{-}$ and $B^{0} \rightarrow \phi\pi^{+}\pi^{-}$ are interesting because they are induced by the  flavor changing neutral current \cite{Buras:1996wn} $ b\rightarrow s\bar{s}s $ and $b\rightarrow d\bar{s}s$ process at the elementary particle level, which is absolutely forbidden at the tree level by the Cabibbo-Kobayashi-Maskawa (CKM) quark-mixing mechanism of the SM \cite{Wang:2018xux}. However, these decays are responsive to the new physics beyond SM because their amplitude are described by the loop (penguin) diagrams \cite{Raidal:2002ph}. After the experimental findings, the rare decay of $B^{0}_{s} \rightarrow \phi\pi^{+}\pi^{-}$ had been studied using the perturbative QCD approach in Ref. \cite{Wang:2018xux}, where the nonperturbative contributions from the resonance $f_0(980)$ are introduced in the distribution amplitudes by the time-like scalar form factor parameterized with the Flatt\'e model. Applying the QCD factorization framework, the three-body decays of $B^0_{(s)} \to \phi \pi^+\pi^-$ are also investigated in Ref. \cite{Estabar:2018ecl}, where the resonant contributions are taken into account for the three-body matrix element in terms of the Breit-Wigner formalism. Besides, also applying the perturbative QCD approach,  the work of \cite{Li:2019xwh} research the direct $CP$ violation in the decay of $B^{0}_{s} \to \rho (\omega) \phi \to \phi\pi^{+}\pi^{-}$ via the $\rho-\omega$ mixing mechanism.

In the present work, aiming at examining the resonant contributions and understanding the reproductions of the $f_{0}(500)$ and $f_{0}(980)$ states  in the final state interactions, we also study the decays of $B^{0}_{s} \rightarrow \phi\pi^{+}\pi^{-}$ and $B^{0} \rightarrow \phi\pi^{+}\pi^{-}$ with the final state interaction approach under the ChUA as done in Refs. \cite{Liang:2014tia,Liang:2015qva,Bayar:2014qha}, where the $\bar{B}^{0}_{s}$ and $\bar{B}^{0}$ mesons decay to $J/\psi$ with $\pi^+ \pi^-$ and the other final states are studies, especially $J/\psi$ with a vector meson considered in Ref. \cite{Bayar:2014qha}. As found in Refs. \cite{Liang:2014tia,Bayar:2014qha}, the $f_{0}(980)$ production is the dominant one in the $\bar{B}^{0}_{s}$ decay where there is indeed no evident signal for the $f_{0}(500)$ state as the experimental findings \cite{LHCb:2012ae,Aaij:2014emv}, whereas the production of the $f_{0}(500)$ resonance is dominant one in the $\bar{B}^{0}$ decay. As already known, the states of $f_{0}(500)$ (or called as $\sigma$ state), $f_{0}(980)$ and $a_{0}(980)$ are dynamically reproduced in the coupled channel interactions via the potentials derived from the lowest order chiral Lagrangian \cite{Gasser:1983yg,Bernard:1995dp} in the work of \cite{Oller:1997ti} taking the chiral dynamics as done in Ref. \cite{Kaiser:1995cy}. Recently, also starting with the ChUA (more applications about the ChUA can be found in recent reviews \cite{Oller:2019opk,MartinezTorres:2020hus,Oller:2020guq,Guo:2020hli}), the work of \cite{Ahmed:2020kmp} researches the different properties of these states $f_{0}(500)$, $f_{0}(980)$ and $a_{0}(980)$ in details, where the couplings, the compositeness, the wave functions and the radii are calculated to reveal the nature of them. 

Indeed, the mixing components for the states of $f_{0}(500)$ and $f_{0}(980)$, which are all mainly decay into the $\pi\pi$ channel, are studied in Ref.  \cite{Agaev:2017cfz} with the QCD sum rule, where more experimental data are required to clarify the strange component in the $f_{0}(500)$ resonance as suggested in Ref. \cite{Agaev:2018sco}. Thus, in the present work, we investigate the properties of the $f_{0}(500)$ and $f_{0}(980)$ states in the productions of the final state interactions on the decay processes of $B^{0}_{s} \rightarrow \phi\pi^{+}\pi^{-}$ and $B^{0} \rightarrow \phi\pi^{+}\pi^{-}$. In the next section, we will briefly introduce the formalism of the final state interactions with the ChUA for these two decay procedures of the $B^{0}_{s}$ and $B^{0}$ mesons. In the following section, we discuss the vector meson productions in the decay procedures. Then, we show the results of the $\pi^+\pi^-$ invariant mass distributions and the branching fractions of some decay channels in the following section. At the end, we make a short conclusion.

\section{The model for scalar meson production}

\begin{figure}
\begin{subfigure}{0.48\textwidth}
  \centering
  \includegraphics[width=1\linewidth]{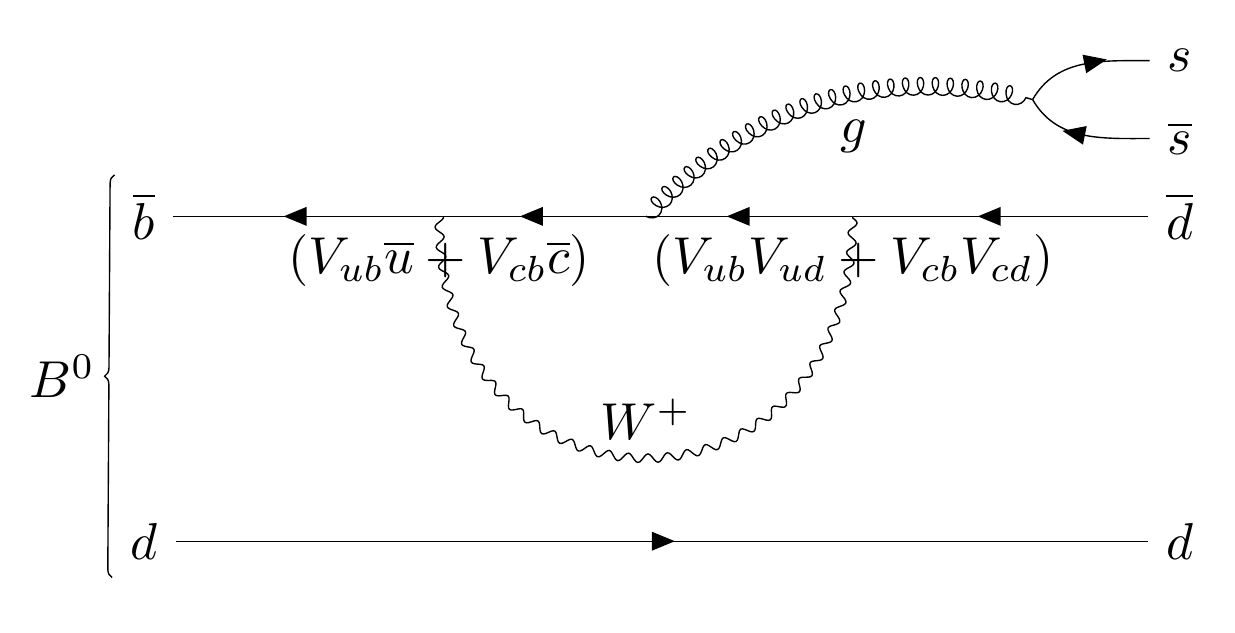}
\caption{\footnotesize $B^{0}$ decays with a $d \bar{d}$ productions.}
\label{fig:fig1a}
\end{subfigure} 
\begin{subfigure}{0.48\textwidth}
  \centering
  \includegraphics[width=1\linewidth]{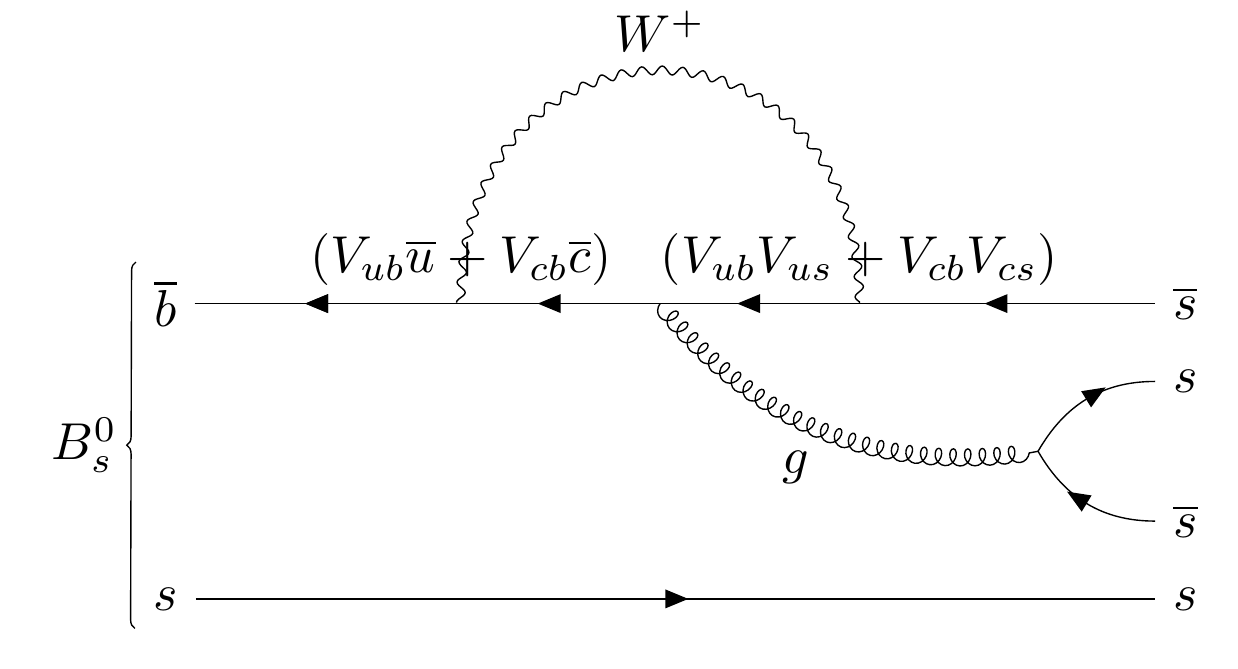}
\caption{\footnotesize  $B^{0}_{s}$ decays with a $s\bar{s}$ productions.}
\label{fig:figb}
\end{subfigure}%
\caption{Feynman diagrams for the decays of  $B^{0}$ and $B^{0}_{s}$ into $\phi$ and a primary $q \bar{q}$ pair. }
\label{fig:fig1}
\end{figure}

\begin{figure}
  \centering
  \includegraphics[width=0.5\linewidth]{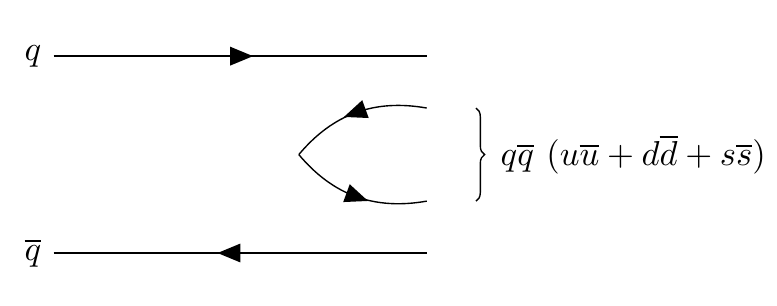}
\caption{Procedure for the hadronization $q\bar{q} \rightarrow q\bar{q}(u\bar{u}+d\bar{d}+s\bar{s})$.}
\label{fig:fig2}
\end{figure}

Following the work of Refs. \cite{Liang:2014tia,Liang:2015qva}, where the decays of $B^0_{(s)} \to J/\psi\pi^{+}\pi^{-}$ are studied, we investigate the analogous ones of $B^0_{(s)} \to \phi\pi^{+}\pi^{-}$ with the vector meson $J/\psi$ replaced by a $\phi$ in the hadron level. But viewing at the dominant weak decay mechanism, the $B^0_{(s)}$ decayed into $J/\psi$ and a $q \bar{q}$ pair in their cases can be easy to fulfil through the tree level $b \to c$ transition. In our cases, it should be proceeded via a gluonic $b \to s$ penguin transition for the $B^0_{(s)}$ decayed into $\phi$ and a $q \bar{q}$ pair, see Fig. \ref{fig:fig1}, and thus, these decay processes are suppressed because these decays are forbidden at the tree level by the Cabibbo-Kobayashi-Maskawa (CKM) quark-mixing mechanism, where we also discuss the suppression effect later. Since the parts of the weak decay mechanism are isolated with a dynamical factor (see our formalism later), in our cases we also apply the final state interaction framework to research the decays of $B^0_{(s)} \to \phi\pi^{+}\pi^{-}$ to focus on the procedure of the $q \bar{q}$ pair hadronized to the final states with the interactions of each other, where this parts of the interactions can be utilized by the coupled channel interactions with the ChUA. Then, we show our formalism in details below. The dominant weak decay mechanism for the $B^{0}$ and $B^{0}_{s}$ decays as depicted in Fig. \ref{fig:fig1} are proceeded as,
\begin{equation}
\begin{aligned} 
B^{0} (\bar{b} d) &\Rightarrow (V_{ub} \bar{u} + V_{cb} \bar{c}) \mathit{W^{+}} d \Rightarrow (V_{ub} \bar{u} g + V_{cb} \bar{c} g) \mathit{W^{+}} d \\
&\Rightarrow (V_{ub} V_{ud} + V_{cb} V_{cd} ) (s\bar{s}) (d\bar{d}) \, ,
\end{aligned} 
\end{equation}
\begin{equation}
\begin{aligned} 
B^{0}_{s}(\bar{b} s) &\Rightarrow (V_{ub} \bar{u} + V_{cb} \bar{c}) \mathit{W^{+}} s \Rightarrow (V_{ub} \bar{u} g + V_{cb} \bar{c} g) \mathit{W^{+}} s \\
&\Rightarrow (V_{ub} V_{us} + V_{cb} V_{cs} ) (s\bar{s}) (s\bar{s})  \, ,
\end{aligned} 
\end{equation}
where $V_{q_1 q_2}$ is the element of the CKM matrix for the transition of the quark $q_1 \to q_2$ (see appendix \ref{ckm} for the details of the CKM matrix), and a $\phi$ with $(s\bar{s})$ and a primary $q \bar{q}$ pair are produced at the end.  To produce the $\pi^{+}\pi^{-}$ mesons alongside with the $\phi$ meson in the final states, the primary $q\bar{q}$ pair must undergo the hadronization. And thus, to achieve this procedure, an additional $q\bar{q}$ pair should be generated from the vacuum to accompany with the primary $q\bar{q}$, written as $u\bar{u}+d\bar{d}+s\bar{s}$, as shown in Fig. \ref{fig:fig2}, where this procedure is formulated as,
\begin{equation}
\begin{aligned} 
B^{0} &\Rightarrow (V_{ub} V_{ud} + V_{cb} V_{cd}) (s\bar{s}\to  \phi) [d\bar{d} \to d\bar{d} \cdot (u\bar{u}+d\bar{d}+s\bar{s}) ] \\
&\Rightarrow  (V_{ub} V_{ud} + V_{cb} V_{cd})  (s\bar{s}\to  \phi) [ M_{22} \to (M \cdot M)_{22} ] \, ,
\end{aligned} 
\end{equation}
\begin{equation}
\begin{aligned} 
B^{0}_{s} &\Rightarrow (V_{ub} V_{us} + V_{cb} V_{cs} )  (s\bar{s}\to  \phi)  [ s\bar{s} \to s\bar{s} \cdot (u\bar{u}+d\bar{d}+s\bar{s}) ] \\
& \Rightarrow (V_{ub} V_{us} + V_{cb} V_{cs} )  (s\bar{s}\to  \phi)  [ M_{33} \to (M \cdot M)_{33} ] \, ,
\end{aligned} 
\end{equation}
with the $q\bar{q}$ matrix element $M$ defined as
\begin{equation}
M=\left(\begin{array}{lll}{u \bar{u}} & {u \bar{d}} & {u \bar{s}} \\ {d \bar{u}} & {d \bar{d}} & {d \bar{s}} \\ {s \bar{u}} & {s \bar{d}} & {s \bar{s}}\end{array}\right) .
\label{eq:matrM}
\end{equation}

\begin{figure}
\begin{subfigure}{1\textwidth}
  \centering
  \includegraphics[width=1\linewidth]{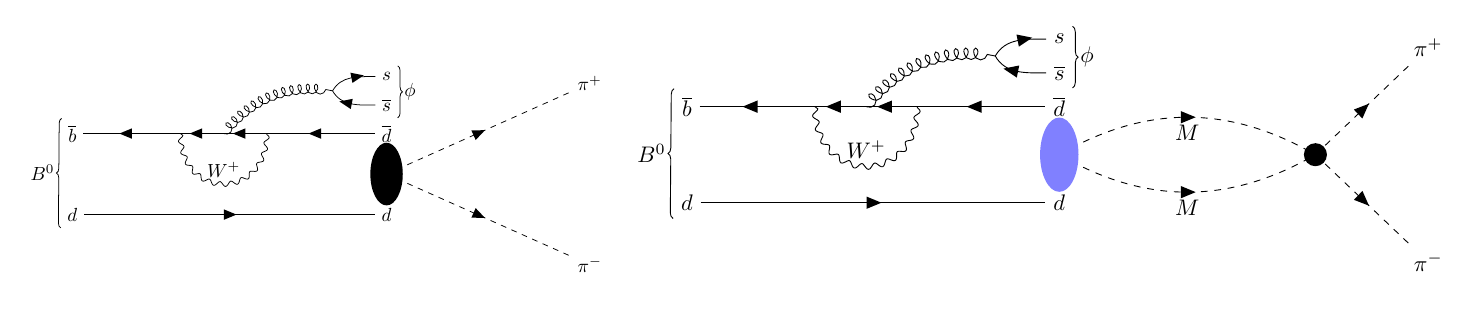}
  \caption{\footnotesize Produced with direct plus rescattering mechanisms in $B^{0}$ decay.}
\end{subfigure}%

\begin{subfigure}{1\textwidth}
  \centering
  \includegraphics[width=0.7\linewidth]{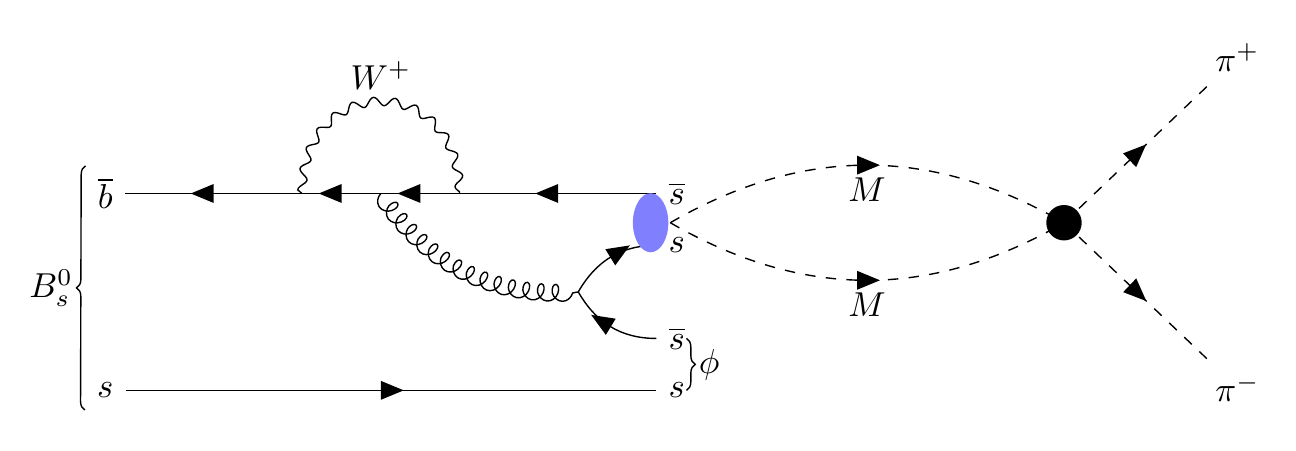}
  \caption{\footnotesize Produced via the rescattering mechanism in $B^{0}_{s}$ decay.}
\end{subfigure}%
\caption{Diagrammatic representation for the $\pi^{+}\pi^{-}$ productions in the final state interactions of $B^{0}$ (a) and $B^{0}_{s}$ (b) decays.}
\label{fig:fig3} 
\end{figure}

Furthermore, we can write the matrix elements of $M$ in terms of the physical mesons, which corresponds to 
\begin{equation}
\Phi=\left(\begin{array}{ccc}{\frac{1}{\sqrt{2}} \pi^{0}+\frac{1}{\sqrt{6}} \eta} & {\pi^{+}} & {K^{+}} \\ {\pi^{-}} & {-\frac{1}{\sqrt{2}} \pi^{0}+\frac{1}{\sqrt{6}} \eta} & {K^{0}} \\ {K^{-}} & {\bar{K}^{0}} & {-\frac{2}{\sqrt{6}} \eta}\end{array}\right),
\label{eq:Mphi}
\end{equation}   
where we take $\eta \equiv \eta_{8}$. With the correspondence between the matrix $M$ and $\Phi$, the hadronization process can be accomplished to the hadron level in terms of two pseudoscalar mesons
\begin{equation}
\begin{aligned}  
d \bar{d} \cdot (u \bar{u}+d \bar{d}+s \bar{s}) & \to (\Phi \cdot \Phi)_{22} =\pi^{+} \pi^{-}+\frac{1}{2} \pi^{0} \pi^{0}-\frac{1}{\sqrt{3}} \pi^{0} \eta+K^{0} \bar{K}^{0}+\frac{1}{6} \eta \eta ,\\ 
s \bar{s} \cdot (u \bar{u}+d \bar{d}+s \bar{s}) & \to (\Phi \cdot \Phi)_{33}=K^{-} K^{+}+K^{0} \bar{K}^{0}+\frac{4}{6} \eta \eta \, ,
\end{aligned}
\label{eq4}
\end{equation} 
where one can see that, there are only $K\bar{K}$ and $\eta\eta$ produced in the $B^{0}_{s}$ decay, which is different from the one of $B^{0}$ decay having the other productions ($\pi\pi$ for example) too. As we have known from the ChUA \cite{Oller:1997ti,Ahmed:2020kmp}, the $f_0(980)$ state is bound by the $K\bar{K}$ component, whereas, the $f_0(500)$ resonance is mainly contributed from the $\pi\pi$ channel. Thus, one can expect these two states will have different contributions for the $B^{0}$ and $B^{0}_{s}$ decays, see our results later. Once the final states are hadronized after the weak decay productions, they also can go to further interactions, as depicted in Fig. \ref{fig:fig3}, where in fact there are three processes taken into account, the $\phi$ emission in the weak decay, the meson pair creation in the $q\bar{q}$ hadronizations and the final state interactions of the hadronic pair, as discussed in details in Ref. \cite{Miyahara:2015cja}. Then, the amplitudes for these final state production and their interaction procedures can be written as
\begin{equation}
\begin{aligned} 
t\left(B^{0} \rightarrow \phi \pi^{+} \pi^{-}\right) =& V_{P} (V_{ub}  V_{ud}  + V_{cb}  V_{cd} )\left(1+G_{\pi^{+} \pi^{-}} T_{\pi^{+} \pi^{-} \rightarrow \pi^{+} \pi^{-}}+ \frac{1}{2} G_{\pi^{0} \pi^{0}} T_{\pi^{0} \pi^{0} \rightarrow \pi^{+} \pi^{-}}\right.\\ 
&\left.+G_{K^{0} \bar{K}^{0}} T_{K^{0} \bar{K}^{0} \rightarrow \pi^{+} \pi^{-}}+ \frac{1}{6} G_{\eta \eta} T_{\eta \eta \rightarrow \pi^{+} \pi^{-}}\right)  \,  , 
\end{aligned}
\label{eq5}
\end{equation}
\begin{equation}
\begin{aligned}  
t\left(B^{0}_{s} \rightarrow \phi \pi^{+} \pi^{-}\right)=& V_{P} (V_{ub}  V_{us}  +V_{cb}  V_{cs} ) \left( G_{K^{+}K^{-}}T_{K^{+}K^{-} \rightarrow \pi^{+} \pi^{-}}+G_{K^{0} \bar{K}^{0}} T_{K^{0} \bar{K}^{0} \rightarrow \pi^{+} \pi^{-}}+ \right.\\ 
& \left. \frac{4}{6} G_{\eta \eta} T_{\eta \eta \rightarrow \pi^{+} \pi^{-}}\right), 
\end{aligned}
\label{eq51}
\end{equation}
where $V_{P}$ \footnote{Note that, we only use the flavor structure of these processes and the remaining dynamical factors are included in $V_{p}$, which is then taken as a constant and independent on $M_{\text{inv}}$ \cite{Li:2012sw}.} is the production vertex factor, which contains all the dynamical factors and is assumed to be universal for these two reactions because of the similar production dynamics and the differences specified by the CKM matrix elements $V_{q_1 q_2}$. 
It is worth to mention that,  in Eqs. \eqref{eq5} and \eqref{eq51}, there is a factor of $2$ in the terms related with the identical particles (such as the $\pi^{0}\pi^{0}$ and $\eta\eta$) because of the two possibilities in the operators of Eq. \eqref{eq4} to create them, which has been cancelled with the factor of $\frac{1}{2}$ in their followed propagators within our normalization schem, see more discussions in Ref. \cite{Liang:2015qva}.
Besides, the scattering amplitude of $T_{ij}$ for the transition of $i \to j$ channel is evaluated by the Bethe-Salpeter equation with the on-shell approximation for the coupled channel interactions,
\begin{equation}
T = [1-VG]^{-1}V ,  \label{eq:BS}
\end{equation}
where the element of the diagonal matrix $G$ is the loop functions of two meson propagators, given by
\begin{equation}
 G _ { i i} ( s ) = i \int \frac { d ^ { 4 } q } { ( 2 \pi ) ^ { 4 } } \frac { 1 } { q ^ { 2 } - m _ { 1 } ^ { 2 } + i \varepsilon } \frac { 1 } { \left( p _ { 1 } + p _ { 2 } - q \right) ^ { 2 } - m _ { 2 } ^ { 2 } + i \varepsilon }  \text{ ,}
 \label{eq:eq6}
\end{equation}
with $p_{1}$ and $p_{2}$ the four-momenta of the two mesons in the certain channel, respectively, having $s=(p_1 + p_2)^2$, and $m_{1}$, $m_{2}$ the corresponding masses for them. Since Eq. \eqref{eq:eq6} is logarithmically divergent, the regularization schemes should be utilized to solve this singular integral, either applying the three-momentum cutoff approach \cite{Oller:1997ti}, where the analytic expression is given by Refs. \cite{Oller:1998hw,Guo:2006fu}, or the dimensional regularization method \cite{Oller:2000fj}. In the present work, we take the cutoff method \cite{Oller:1997ti} for Eq. \eqref{eq:eq6},
\begin{equation}
G _ { ii } ( s ) = \int _ { 0 } ^ { q _ { \max } } \frac { q ^ { 2 } d q } { ( 2 \pi ) ^ { 2 } } \frac { \omega _ { 1 } + \omega _ { 2 } } { \omega _ { 1 } \omega _ { 2 } \left[ s - \left( \omega _ { 1 } + \omega _ { 2 } \right) ^ { 2 } + i \varepsilon \right] }   \text{  ,}
\end{equation}
with $q=|\vec{q}\,|$ and $ \omega _ { i } = ( \vec { q } ^ { \:2 } + m _ { i } ^ { 2 } ) ^ { 1 / 2 }$, where the free parameter of the cutoff $q_{max}$ is chosen as $600$ MeV for the case of including $\eta\eta$ channel \cite{Liang:2014tia} and $931$ MeV for the one of excluding $\eta\eta$ channel \cite{Xiao:2019lrj}. Furthermore, the matrix $V$ is constructed by the scattering potentials of each coupled channel, where the elements for the $\pi\pi$ and $ K\bar{K}$ channels are taken from Ref. \cite{Oller:1997ti} and the one for the $\eta\eta$ channel from Ref. \cite{Gamermann:2006nm}. Thus, after projecting the potential to $S$-wave, the elements of $V$ matrix, $V_{ij}$,  are given by
\begin{equation}
\begin{aligned}
&V_{11}=-\frac{1}{2 f^{2}} s, \quad V_{12}=-\frac{1}{\sqrt{2} f^{2}}\left(s-m_{\pi}^{2}\right), \quad V_{13}=-\frac{1}{4 f^{2}} s ,\\
&V_{14}=-\frac{1}{4 f^{2}} s, \quad V_{15}=-\frac{1}{3 \sqrt{2} f^{2}} m_{\pi}^{2}, \quad V_{22}=-\frac{1}{2 f^{2}} m_{\pi}^{2} ,\\
&V_{23}=-\frac{1}{4 \sqrt{2} f^{2}} s, \quad V_{24}=-\frac{1}{4 \sqrt{2} f^{2}} s, \quad V_{25}=-\frac{1}{6 f^{2}} m_{\pi}^{2} ,\\
&V_{33}=-\frac{1}{2 f^{2}} s, \quad V_{34}=-\frac{1}{4 f^{2}} s ,\\
&V_{35}=-\frac{1}{12 \sqrt{2} f^{2}}\left(9 s-6 m_{\eta}^{2}-2 m_{\pi}^{2}\right), \quad V_{44}=-\frac{1}{2 f^{2}} s ,\\
&V_{45}=-\frac{1}{12 \sqrt{2} f^{2}}\left(9 s-6 m_{\eta}^{2}-2 m_{\pi}^{2}\right) ,\\
&V_{55}=-\frac{1}{18 f^{2}}\left(16 m_{K}^{2}-7 m_{\pi}^{2}\right),
\end{aligned}
\end{equation} 
where the indices 1 to 5 denote the five coupled channels of $\pi^{+}\pi^{-}$, $\pi^{0}\pi^{0}$, $K^{+}K^{-}$, $K^{0}\bar{K}^{0}$, and $\eta\eta$, respectively, and $f$ is the pion decay constant, taken as $93$ MeV \cite{Oller:1997ti}. Note that, a normalization factor $\frac{1}{\sqrt{2}}$ has been taken into account in the corresponding channels with the identical states of $\pi^0\pi^0$ and $\eta\eta$, and thus, there is no such a factor in the corresponding loop functions in the $G$ matrix, see Eq. \eqref{eq:BS}.

In order to analysis the $\pi\pi$ invariant mass distributions as given in Ref. \cite{Aaij:2016qnm}, we need to evaluate the differential decay width $\frac{d\Gamma}{dM_{\text{inv}}}$ in terms of the $\pi^{+}\pi^{-}$ invariant mass $M_{\text{inv}}$. Before doing that, one need to know the partial waves for the final states. If the hadronization parts of $q\bar{q} (\to\pi\pi)$ is in $S$-wave (see for $P$-wave in the next section), which lead to its $J^P =L^{(-1)^L} = 0^+$, thus, the primary decay procedure is a $0^{-} \rightarrow 1^{-} \,+\,  0^{+}$ transition. Therefore, the angular momentum conservation requires a $P$-wave $L'=1$ for the outgoing vector meson $\phi$, and then, there will be a term of $p_\phi \cos\theta$ contributed to the decay amplitude. Thus, we have finally
\begin{equation}
\frac{d \Gamma}{d M_{\text{inv}}}=\frac{1}{(2 \pi)^{3}} \frac{1}{8 M_{B^0_{(s)}}^{2}} \frac{2}{3} p_{\phi}^{3} \tilde{p}_{\pi}  \bar{\sum} \sum \left| t_{B_{(s)}^{0} \rightarrow \phi \pi^{+} \pi^{-}}\right|^{2},
\label{eq10}
\end{equation} 
where the factor $\frac{2}{3}$ comes from the integral of $\cos^{2}\theta$. Note that, when we fit the $\pi\pi$ invariant mass distributions, we take $\frac{d \Gamma}{d M_{\text{inv}}} \to C \frac{d \Gamma}{d M_{\text{inv}}}$ with an arbitrary constant $C$ to match the events of the experimental data, see our results later. Besides, $p_{\phi}$ is the $\phi$ momentum in the rest frame of  the decaying $B^0_{(s)}$ meson, and $\tilde{p}_{\pi}$ the pion momentum in the rest frame of the $\pi^{+}\pi^{-}$ system, which are given by
\begin{equation}
\begin{aligned}
&p_{\phi}=\frac{\lambda^{1 / 2}\left(M_{B^0_{(s)}}^{2}, M_{\phi}^{2}, M_{\text{inv}}^{2}\right)}{2 M_{B_{(s)}}},\\
&\tilde{p}_{\pi}=\frac{\lambda^{1 / 2}\left(M_{\text{inv}}^{2}, m_{\pi}^{2}, m_{\pi}^{2}\right)}{2 M_{\text {inv }}},
\end{aligned}
\end{equation}
with the usual K\"allen triangle function $\lambda(a, b, c) = a^{2} + b^{2} + c^{2} - 2(ab + ac + bc)$.

\section{The model for vector meson production}

\begin{figure}
  \centering
  \includegraphics[width=0.5\linewidth]{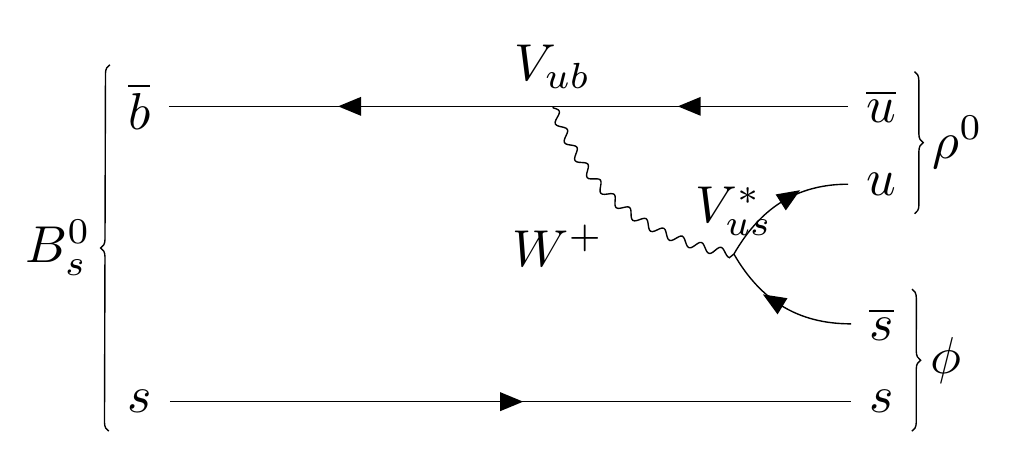}
\caption{Diagram of $B^{0}_{s}$ decay into $\phi $ and $\rho^{0}$ mesons. }
\label{fig:fig4}
\end{figure} 

As discussed above, the hadronization parts $q\bar{q} (\to\pi\pi)$ can also be in $P$-wave, and thus, the quantum numbers of this parts is $J^P =L^{(-1)^L} = 1^-$, which corresponds to the vector meson production. Therefore, for the transition $0^- \rightarrow 1^- \,+\,  1^-$, the created vector mesons will be in the partial waves of $L=0,2$ to preserve the angular momentum conservation. But, as done in Ref. \cite{Bayar:2014qha}, we also take $L=0$ for simplification, and thus, there is no term of $p_{\phi} \cos\theta$ present in the amplitudes. Taking the vector meson $\rho^0$ (except for $\phi$) production for example (the others discussed later), see Fig. \ref{fig:fig4} for the  $B^{0}_{s} \to \phi \rho^{0}$ decay, the amplitude associated to this decay is given by
\begin{equation}
t_{B^{0}_{s} \rightarrow \phi\rho^{0}} = \frac{1}{\sqrt{2}} \tilde{V}_{P} V_{ub}  V_{us}^{*}  \textbf{ },
\end{equation}
where the prefactor $\frac{1}{\sqrt{2}}$ is the $u\bar{u}$ component in $\rho^{0}$, and $\tilde{V}_{P}$ the production vertex factor which contains all the dynamical factors and analogous to the one $V_P$ above but different. In general, the width for the vector meson ($V$) decay $B^0_{(s)} \to \phi V$ is given by
\begin{equation}
\Gamma_{B^{0}_{(s)} \rightarrow \phi V}=\frac{1}{8 \pi} \frac{1}{m_{B_{(s)}^{0}}^{2}}\left|t_{B^{0}_{(s)} \rightarrow \phi V}\right|^{2} p_{\phi}.
\label{eq13}
\end{equation} 

Next, since the produced vector meson $\rho^0$ can be easily decay into $\pi^+\pi^-$, its contributions to the $\pi^+\pi^-$ invariant mass distributions in the $B^{0}_{s}$ decay can be obtained by means of the spectral function as done in Refs. \cite{Bayar:2014qha,Liang:2014ama,Wang:2020pem},
\begin{equation}
\frac{d \Gamma_{B^{0}_{s} \rightarrow \phi \rho^{0}}}{d M_{\text{inv}}\left(\pi^{+} \pi^{-}\right)}=- \frac{1}{\pi} 2 m_{\rho} \operatorname{Im} \frac{1}{M_{\text{inv}}^{2}-m_{\rho}^{2}+i m_{\rho} \tilde{\Gamma}_{\rho}\left(M_{\text{inv}}\right)} \Gamma_{B^{0}_{s} \rightarrow \phi \rho^{0}},
\label{eq14}
\end{equation}
where $\tilde{\Gamma}_{\rho}\left(M_{\text{inv}}\right)$ is the energy dependent decay width of $\rho^{0}$ into two pions, which is given by the parameterization,
\begin{equation}
\begin{aligned}
&\tilde{\Gamma}_{\rho}\left(M_{\text{inv}}\right)=\Gamma_{\rho}\textbf{ } \left(\frac{p_{\pi}^{\text {off }}}{p_{\pi}^{\text {on }}}\right)^{3},\\
&p_{\pi}^{\mathrm{off}}=\frac{\lambda^{1 / 2}\left(M_{\text{inv}}^{2}, m_{\pi}^{2}, m_{\pi}^{2}\right)}{2 M_{\text{inv}}} \theta\left(M_{\text{inv}}-2 m_{\pi}\right),\\
&p_{\pi}^{\mathrm{on}}=\frac{\lambda^{1 / 2}\left(m_{\rho}^{2}, m_{\pi}^{2}, m_{\pi}^{2}\right)}{2 m_{\rho}},
\end{aligned}
\end{equation}
with $p_{\pi}^{\mathrm{on}}(p_{\pi}^{\mathrm{off}})$ the pion on-shell (off-shell) three momentum in the rest frame of the $\rho^0$ decay, $\Gamma_{\rho}$  the total $\rho^0$ decay width, taking as $\Gamma_{\rho} = 149.1$ MeV \cite{pdg2018}, and the step function of $\theta\left(M_{\text{inv}}-2 m_{\pi}\right)$.

\begin{figure}
  \centering
  \includegraphics[width=0.45\linewidth]{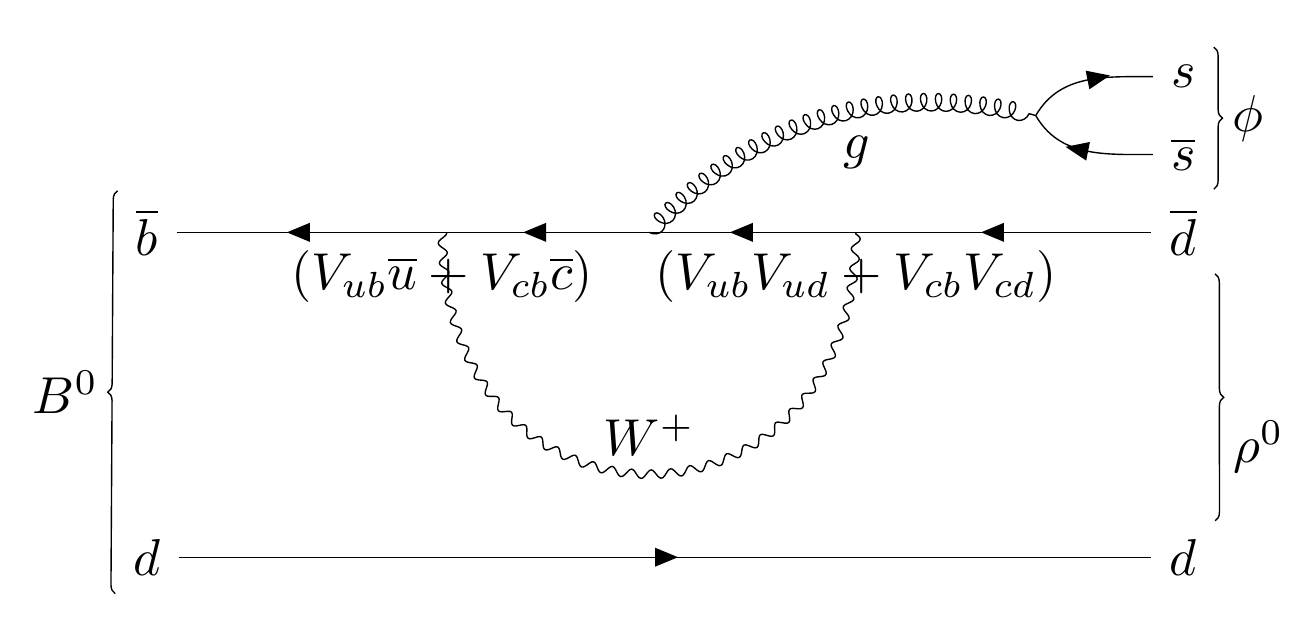}
  \includegraphics[width=0.45\linewidth]{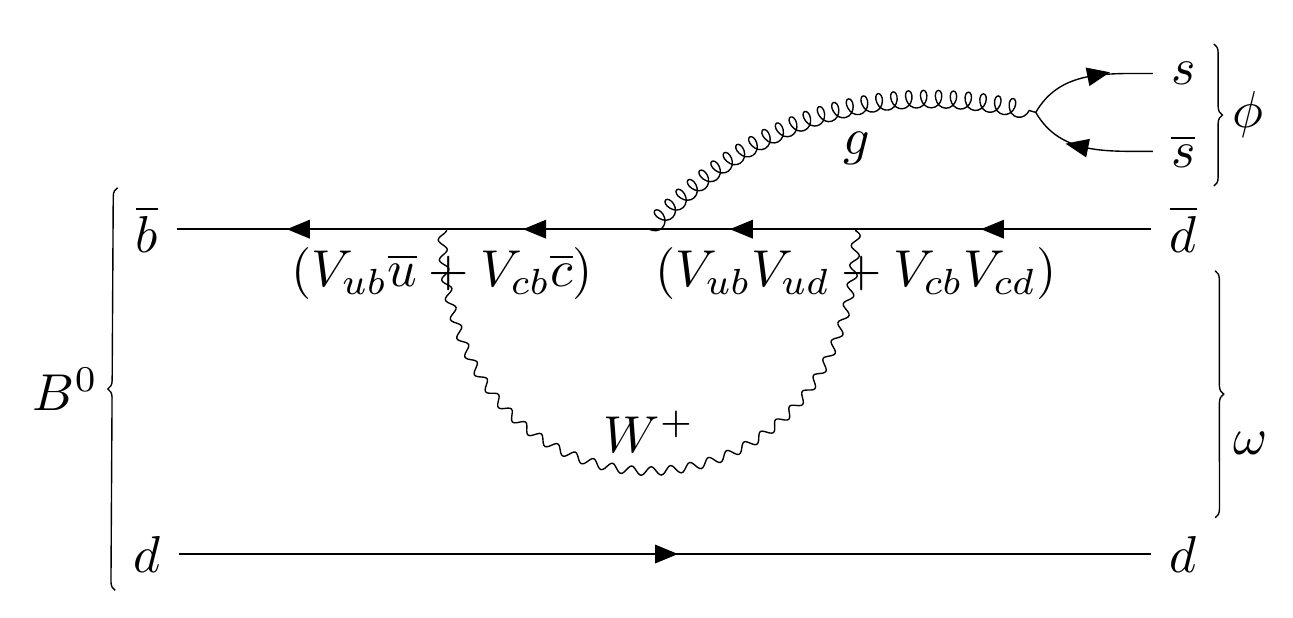}  
  \includegraphics[width=0.45\linewidth]{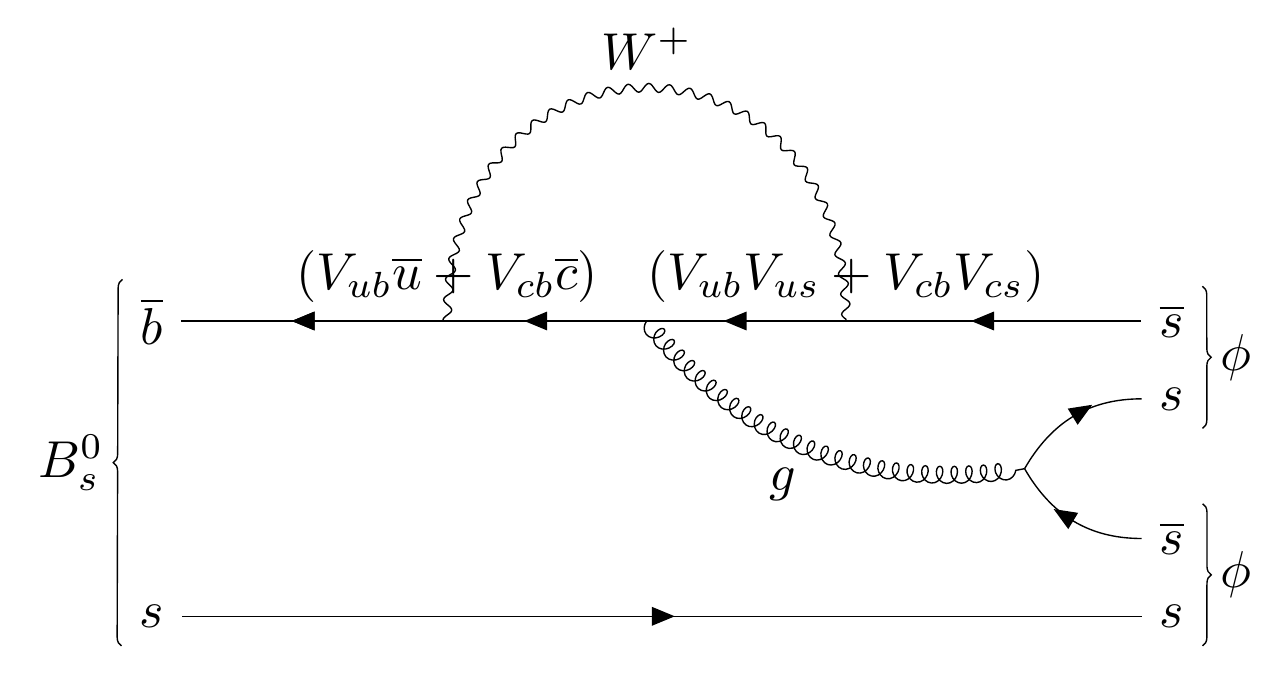}
  \includegraphics[width=0.45\linewidth]{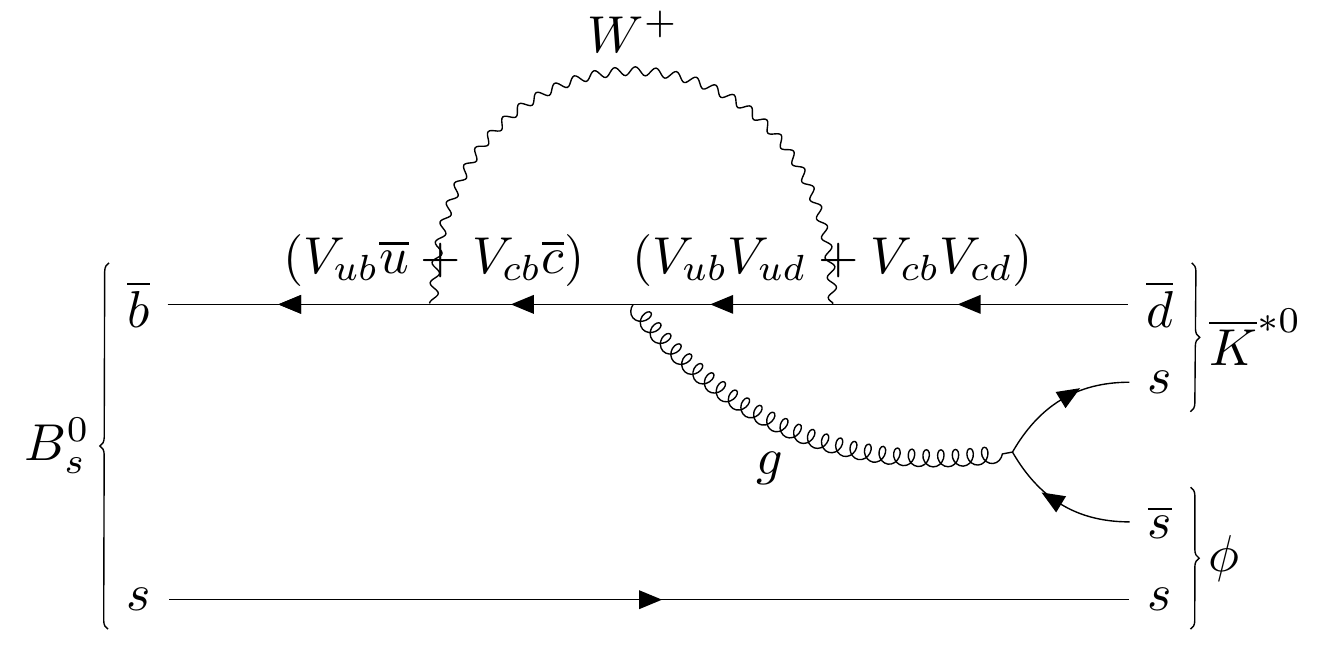}
\caption{Feynman diagrams for $B_{0}$ and $B^{0}_{s}$ decays into $\phi$ and other vector mesons.}
\label{fig:fig41}
\end{figure}

Moreover, we can carry on the investigation of the other vector meson productions, as shown in Fig. \ref{fig:fig41},  where none of these decay processes is allowed at the tree level, see the depression results later, and different from the one of Fig. \ref{fig:fig4}. The corresponding amplitudes of the decay diagrams in Fig \ref{fig:fig41} are written as,
\begin{equation}
\begin{array}{lll}
t_{B^{0} \rightarrow \phi \rho^{0}}=-\frac{1}{\sqrt{2}} \tilde{V}_{P}^{\prime} (V_{ub}  V_{u d}  +V_{c b}  V_{c d} ), \textbf{   } & t_{B^{0} \rightarrow \phi \omega}=\frac{1}{\sqrt{2}} \tilde{V}_{P}^{\prime} (V_{ub}  V_{u d}  + V_{c b}  V_{c d} ),\textbf{   }  \\ t_{B^{0}_{s} \rightarrow \phi \phi}=2 \tilde{V}_{P}^{\prime} (V_{ub}  V_{u s}  +V_{c b}  V_{c s}  ),\textbf{   }
& t_{B^{0}_{s} \rightarrow \phi \bar{ K}^{* 0}}=2 \tilde{V}_{P}^{\prime} (V_{ub}  V_{u d}  +V_{c b}  V_{c d} ),
\end{array}
\end{equation}
where the factor $-\frac{1}{\sqrt{2}}$ is the $d\bar{d}$ component in $\rho^{0}$ whereas $\frac{1}{\sqrt{2}}$ in $\omega$, $\tilde{V}_{P}^{\prime}$ another vertex factor for these hadronization procedures. Note that, one can see an extra factor of two in the two $B^{0}_{s}$ decay modes, $\phi \phi$ and $\phi \bar{ K}^{* 0}$, because there are two possibilities  to create the $\phi$, one by the internal gluon as shown in the lower parts of Fig. \ref{fig:fig41} and the other one by the external gluon analogous to the case of $B^0$ decay in the upper parts of Fig. \ref{fig:fig41}, which are different from the cases of $\Lambda_c$ decay with the internal or external $W$ boson exchange as discussed in Refs. \cite{Wang:2020pem,Li:2020fqp,Xie:2016evi}, where the internal and external $W$ emission mechanism can also be discussed in the recent works on the reactions of $D^+\to \pi^+\pi^0\eta$ \cite{Duan:2020vye} and $D^0\to K^-\pi^+\eta$ \cite{Toledo:2020zxj}. Then, one can calculate the decay widths for these decay modes with the vector productions using Eq. \eqref{eq13}, and thus, the decay ratios for them, see our results in the next section.

\section{Results}

\begin{figure}
  \centering
 \includegraphics[width=0.6\linewidth]{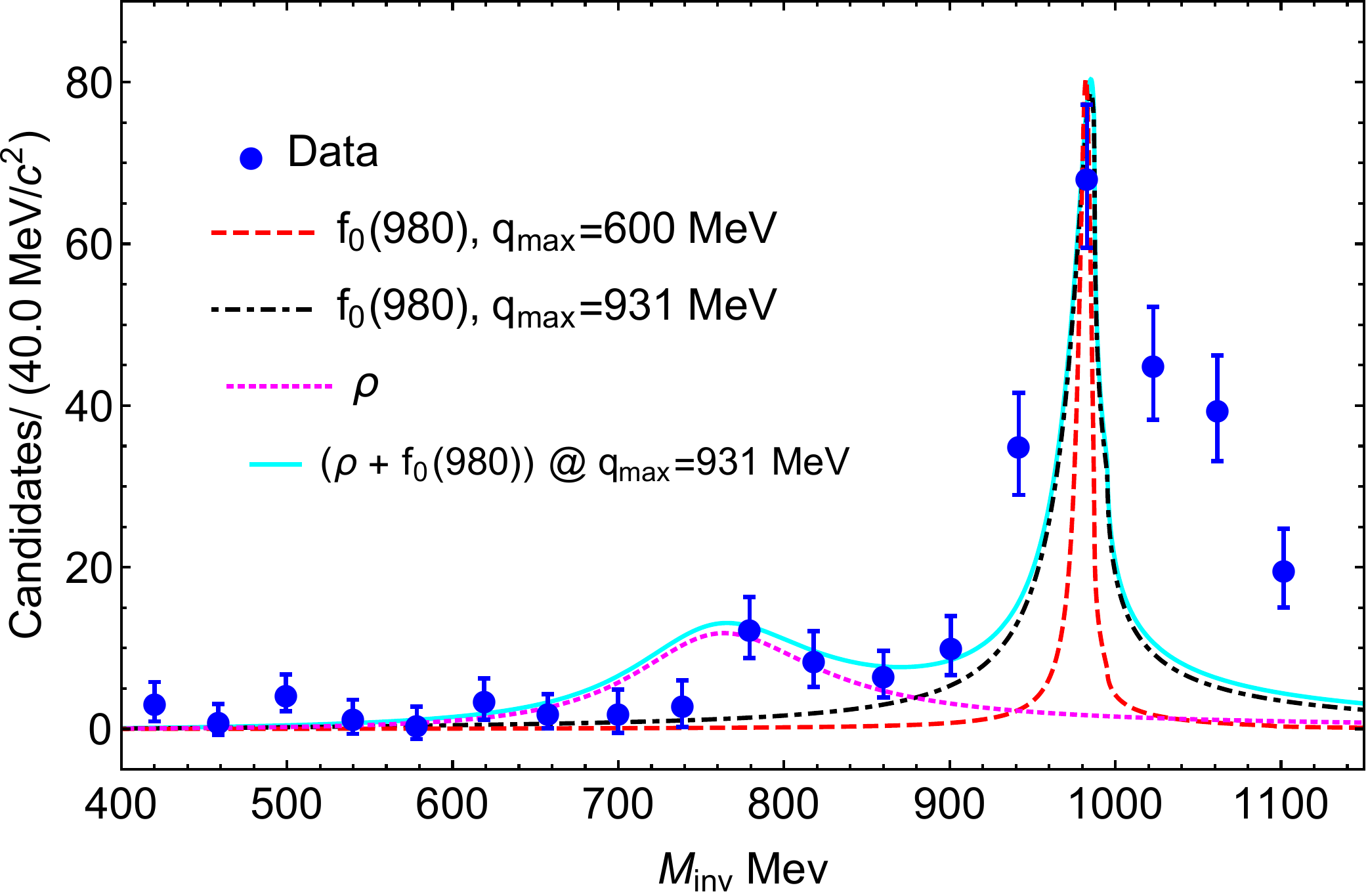} 
\caption{$\pi^{+}\pi^{-}$ invariant mass distributions of the $B^{0}_{s}\rightarrow \phi\pi^{+}\pi^{-}$ decay, where we plot $\frac{C \times 10^{-9}}{\Gamma_{B_{s}}}  \frac{d \Gamma}{d M_{\text{inv}}}$.
The dash (red) line corresponds to the $f_{0}(980)$ resonance contributions with the coupled channel of $\eta\eta$ (normalization constant $C=1.22$), the dash-dot (black) line without the $\eta\eta$ channel ($C=3.70$), and the dot (magenta) line corresponds to the $\rho$ meson contribution ($C=10.0$), and the solid (cyan) line represents the sum of the contributions of two states $f_{0}(980)$ and $\rho$. Data is taken from Ref. \cite{Aaij:2016qnm}.}
\label{fig:fig5}
\end{figure}

\begin{figure}
  \centering
 \includegraphics[width=0.6\linewidth]{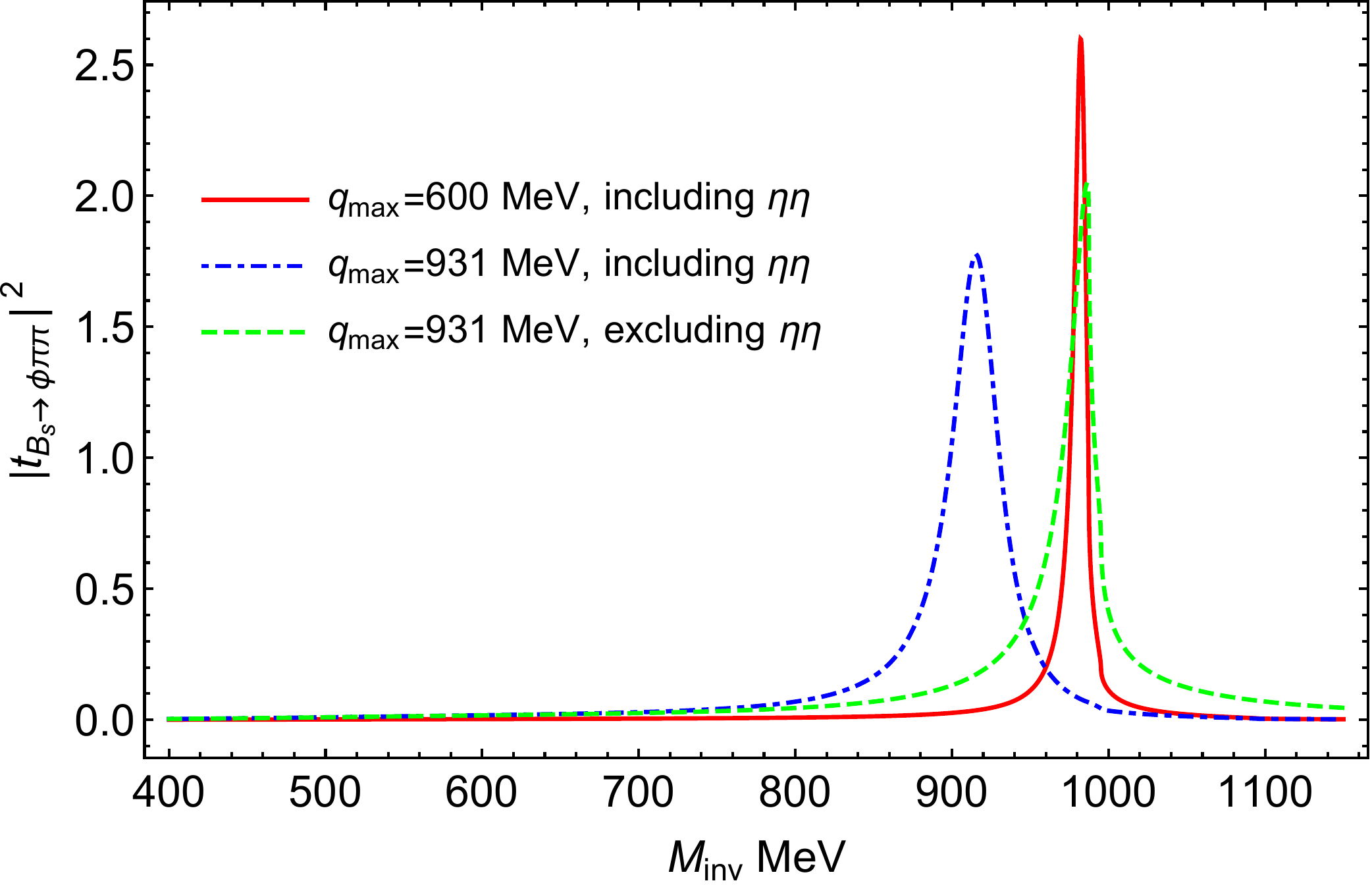} 
\caption{Modulus square of the scattering amplitude of $t_{B^{0}_{s}\rightarrow \phi \pi^{+}\pi^{-}}$, see Eq. \eqref{eq51}, where we only plot the last parts without the previous factors of the vertex $V_P$ and the CKM elements, for the case of including the $\eta\eta$ channel with the cutoff $q_{max}= 600$ MeV (the solid, red line), 931 MeV (the dash-dot, blue line) and excluding the $\eta\eta$ channel with the cutoff $q_{max}= 931$ MeV (the dash, green line). }
\label{fig:fig51}
\end{figure}

As discussed in the introduction, the rare decays of $B^{0}_{s} \rightarrow \phi\pi^{+}\pi^{-}$ and $B^{0} \rightarrow \phi\pi^{+}\pi^{-}$ has been reported by the LHCb collaboration \cite{Aaij:2016qnm}, where the $\pi\pi$ invariant mass distributions and some related branching fractions are obtained. To look for the resonant contributions in the energy region lower than about 1 GeV, we show the results of the $\pi\pi$ invariant mass distributions for the decay of $B^{0}_{s}\rightarrow \phi \pi^{+} \pi^{-}$ in Fig. \ref{fig:fig5}, which is only plotted up to 1.1 GeV within the effective energy range of the ChUA for the meson-meson interactions \cite{Oller:1997ti}. From Fig. \ref{fig:fig5}, one can see that the dominant contributions are the $f_0(980)$ resonant around the region 1 GeV, which is consistent with the one fitted with Flatt\'e model in the experimental results of Ref. \cite{Aaij:2016qnm}. As discussed in the formalism above, there are some theoretical uncertainties for the coupled channel interactions with \cite{Gamermann:2006nm} or without \cite{Oller:1997ti} the $\eta\eta$ channel, see the results of the dash (red, with $q_{max} = 600$ MeV) line and the dash-dot (black, with $q_{max} = 931$ MeV) line in Fig. \ref{fig:fig5}, respectively,  where one can see that the line shape of the $f_0(980)$ state is more narrow when the contribution of the $\eta\eta$ channel is taken into account. Since the threshold of the $\eta\eta$ channel is not so far above the $f_0(980)$, it has nontrivial effects, which will give some uncertainties to the branching ratios, see our results later. Indeed, when the $\eta\eta$ channel is considered, the pole for the $f_0(980)$ state becomes smaller, see the dash-dot (blue) line and the dash (green) line of Fig. \ref{fig:fig51}. But, as found in Ref. \cite{Ahmed:2020kmp}, for the bound state of the $K\bar{K}$ channel, one should decrease the cutoff to move the pole of the $f_0(980)$ state to higher energy when the $\eta\eta$ channel is included, which will lead to the width of the pole decrease, see the solid (red) line of Fig. \ref{fig:fig51} and more discussions in Ref. \cite{Ahmed:2020kmp}. This is why the peak of the $f_0(980)$ state become narrow when we add the coupled channel of $\eta\eta$, which is different from the interference effects in the case of narrow $\sigma$ state in the  $J/\psi \to p \bar{p} \pi^+\pi^-$ decay \cite{Li:2003zi} and the $J/\psi \to \omega\pi\pi$ decay \cite{Roca:2004uc}. Furthermore, the $\eta\eta$ channel was also taken into account in the investigation of the $D^+\to K^-K^+K^+$ decay in a recent work of \cite{Roca:2020lyi}, where the $N/D$ method were used for the scattering amplitudes as done in Ref. \cite{Oller:1998zr} to extend  the applicability range higher than 1 GeV and they found that the dominant contributions were the one of $a_0(980)$ near the $K\bar{K}$ threshold.

\begin{figure}
  \centering
 \includegraphics[width=0.6\linewidth]{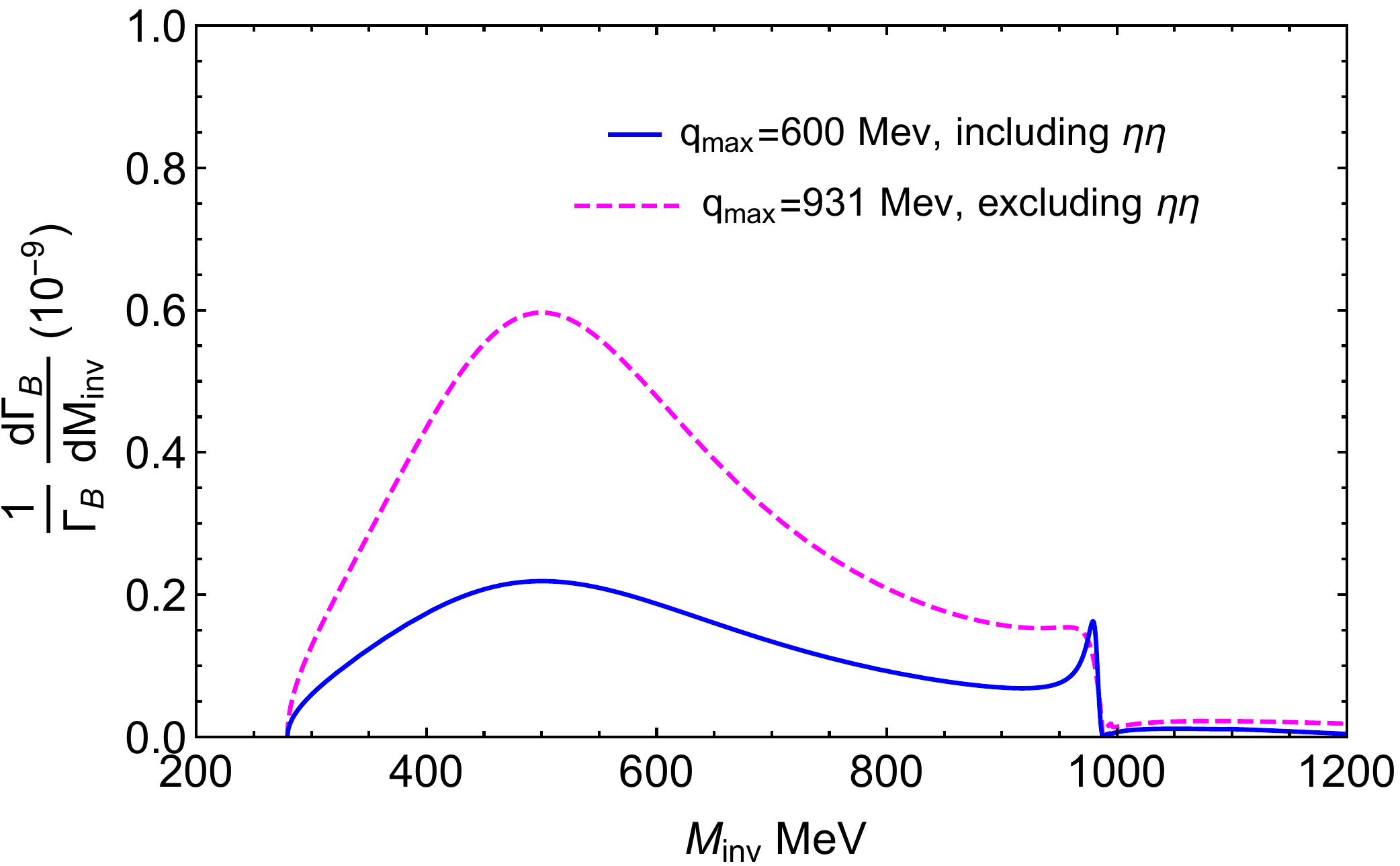}
\caption{$\pi^{+}\pi^{-}$ invariant mass distributions of the $B^{0}\rightarrow \phi\pi^{+}\pi^{-}$ decay for with the $\eta\eta$ channel (blue) and without the $\eta\eta$ channel (magenta).}
\label{fig:fig6}
\end{figure}

In the last section, we also take into account the contributions from the vector meson productions when the final states of $\pi^+\pi^-$ are in $P$-wave, see the results of the dot (magenta) line in Fig. \ref{fig:fig5}, which is the contributions of the $\rho$ meson, as commented in the experiment \cite{Aaij:2016qnm}. From the results of the solid (cyan) line in Fig. \ref{fig:fig5}, our results of the sum of two resonances contributions of $f_0(980)$ and $\rho$ describe the experimental data up to 1 GeV well. But, as found in the experiment \cite{Aaij:2016qnm}, there is no signal for the $f_{0}(500)$ resonance. Indeed, in our formalism there is no such contributions, see Eq. \eqref{eq51}, since the the $f_{0}(500)$ state appears in the amplitude of $T_{\pi^{+} \pi^{-} \rightarrow \pi^{+} \pi^{-}}$, which is analogous to the case of $B^{0}_{s} \rightarrow J/\psi \pi^{+} \pi ^{-}$ decay \cite{Liang:2014tia,Liang:2015qva,Bayar:2014qha} as shown in our reproduced results in the appendices. Conversely, this is not the case for the $B^{0}\rightarrow \phi\pi^{+}\pi^{-}$ decay, see our predicted results of the S-wave $\pi^{+}\pi^{-}$ mass distribution for this decay in Fig. \ref{fig:fig6}, where one can see that the contributions from the broad peak of the $f_{0}(500)$ state above the $\pi \pi $ threshold is more significant than the one of the narrow and small peak near the $K\bar{K}$ threshold corresponded to the $f_{0}(980)$ resonance. There are also some uncertainties for the effects of the $\eta\eta$ channel as shown in Fig. \ref{fig:fig6}, where the contributions of the the $f_{0}(500)$ state is more stronger when the $\eta\eta$ channel is not considered, see the dash (magenta) line of Fig. \ref{fig:fig6}. Note that, there some uncertainties from the values of the CKM matrix elements, taking the Wolfenstein parameterization or the absolute values, see Eqs. \eqref{eq:ckm1} and \eqref{eq:ckm2}, especially in the case of the $B^{0}\rightarrow \phi\pi^{+}\pi^{-}$ decay, see Eq. \eqref{eq5}. But, our main results are used the ones of the Wolfenstein parameterization.

Due to the production vertex $V_{p}$ unknown in our formalism, see the discussion after Eq. \eqref{eq51}, for the predictions of $B^{0}\rightarrow \phi\pi^{+}\pi^{-}$, we need to determine it from the decay of $B^{0}_{s}\rightarrow \phi\pi^{+}\pi^{-}$.  For the case of considering the $\eta\eta$ channel, using Eq. \eqref{eq10} we have
\begin{equation} 
\text{Br}(B^{0}_{s} \rightarrow \phi f_{0}(980))= \frac{\Gamma_{B^{0}_{s} \rightarrow \phi f_{0}(980)}}{\Gamma_{B_{s}}}= \frac{\int_{2m_{\pi}}^{1200}\frac{d\Gamma_{B^{0}_{s} \rightarrow \phi f_{0}(980)}}{dM_{inv}}dM_{inv}}{\Gamma_{B_{s}}}  \\ = \frac{V_{p}^{2}}{\Gamma_{B_{s}}} \times 32.95 \textbf{ },
\label{eq:32}
\end{equation} 
whereas, for the case of without the $\eta\eta$ channel, 
 \begin{equation} 
 \text{Br}(B^{0}_{s} \rightarrow \phi f_{0}(980))= \frac{\Gamma_{B^{0}_{s} \rightarrow \phi f_{0}(980)}}{\Gamma_{B_{s}}}= \frac{\int_{2m_{\pi}}^{1200}\frac{d\Gamma_{B^{0}_{s} \rightarrow \phi f_{0}(980)}}{dM_{inv}}dM_{inv}}{\Gamma_{B_{s}}}  \\ = \frac{V_{p}^{2}}{\Gamma_{B_{s}}} \times 51.18 \textbf{ }.
\label{eq:33}
\end{equation}
And then, using the measured branching fraction of $\text{Br}(B^{0}_{s} \rightarrow \phi f_{0}(980))= (1.12 \pm 0.21) \times 10^{-6}$ \cite{pdg2018}, we can obtain $ \frac{V_{p}^{2}}{\Gamma_{B_{s}}}= (3.40 \pm 0.64) \times 10^{-8}$ and $\frac{V_{p}^{2}}{\Gamma_{B_{s}}}= (2.19 \pm 0.41) \times 10^{-8}$ for the two cases, respectively. The uncertainties presented here are estimated from the errors of the experimental branching ratio. 
Thus, we can predict the branching ratios of the decays $B^{0}\rightarrow \phi f_{0}(980)\rightarrow \phi\pi^{+}\pi^{-}$ and $B^{0}\rightarrow \phi f_{0}(500)\rightarrow \phi\pi^{+}\pi^{-}$ using the determined values of $ \frac{V_{p}^{2}}{\Gamma_{B_{s}}}$ from Eqs. \eqref{eq:32} and \eqref{eq:33} with 
\begin{equation} 
\text{Br}(B^{0} \rightarrow \phi f_{0}(980))= \frac{\Gamma_{B^{0} \rightarrow \phi f_{0}(980)}}{\Gamma_{B}}= \frac{\int_{900}^{1200}\frac{d\Gamma_{B^{0} \rightarrow \phi f_{0}(980)}}{dM_{inv}}dM_{inv}}{\Gamma_{B}}  ,
\label{eq:34}
\end{equation} 
\begin{equation} 
\text{Br}(B^{0} \rightarrow \phi f_{0}(500))= \frac{\Gamma_{B^{0} \rightarrow \phi f_{0}(500)}}{\Gamma_{B}}= \frac{\int_{2m_{\pi}}^{900}\frac{d\Gamma_{B^{0} \rightarrow \phi f_{0}(500)}}{dM_{inv}}dM_{inv}}{\Gamma_{B}}  ,
\label{eq:35}
\end{equation}
where the predicted results are shown in Table \ref{tab:tab1}. Note that, for the results in Table \ref{tab:tab1}, we have considered two uncertainties. The first one is estimated from the experimental error of the branching ratio used in determining the vertex factor, and the second one comes from the limits of the integration of Eqs. \eqref{eq:34} and \eqref{eq:35}, since there is some uncertainties in overlap region for the contribution of $f_{0}(500)$ and $f_{0}(980)$ as shown in Fig. \ref{fig:fig6}. For the central value, we has chosen $900$ MeV for the cutting point of the contributions between the $f_{0}(500)$ state and the $f_{0}(980)$ state, see Eqs. \eqref{eq:34} and \eqref{eq:35}. To estimate the uncertainty, we changed the central value by $\pm 50$ MeV.

\begin{table}
\renewcommand{\arraystretch}{1.7}
     \setlength{\tabcolsep}{0.2cm}
\center
\caption{Predicted branching ratios of $B^{0}\rightarrow \phi f_{0}(980)$ and $B^{0}\rightarrow \phi f_{0}(500)$.}
\resizebox{0.8\textwidth}{!}{\begin{tabular}{|c|c|c|c|}
\hline
Branching ratios & Without $\eta\eta$ channel  & With $\eta\eta$ channel  & Exp.  \\ \hline 
$\text{Br}(B^{0}\rightarrow \phi f_{0}(980))$  &  $(4.69 \pm 0.88 \pm  _{-1.55}^{+3.96}) \times 10^{-9} $   &  $ (7.37 \pm 1.38 _{-2.11}^{+4.61}) \times 10^{-9}$   &  $< 3.8 \times 10^{-7} $  \\ \hline  
$\text{Br}(B^{0}\rightarrow \phi f_{0}(500))$  &    $(6.20 \pm 1.16  _{-0.21}^{+0.24})\times 10^{-8}$      & $(7.17 \pm 1.35_{-0.27}^{+0.31} )\times 10^{-8}$  & -  \\ \hline
\end{tabular}}
\label{tab:tab1}
\end{table}

In addition, analogously we can also make some predictions for the ratios between different final states of $B^{0}_{s}$ and $B^{0}$ decays. In the present work, we study the suppressed decays of the $B^{0}_{s}\rightarrow \phi \pi^{+} \pi^{-}$ and $B^{0}\rightarrow \phi \pi^{+} \pi^{-}$ comparing to the Cabibbo allowed ones of the $B^{0}_{s} \rightarrow J/\psi \pi^{+} \pi^{-}$ and $B^{0}\rightarrow J/\psi  \pi^{+} \pi^{-}$, see Refs. \cite{Liang:2014tia,Liang:2015qva}, which are reproduced in details \cite{Liang:2014tia} in Appendix \ref{section:app1}. Thus, we can predict the ratios between all the other channels relevant to these decays, based on the experimental results of the ratio to get the relation of the vertex factor, given by
\begin{equation}
\frac{\text{Br}(B^{0}_{s} \rightarrow \phi f_{0}(980))}{\text{Br}(B^{0}_{s}\rightarrow J/\psi f_{0}(980))}=(8.75 \pm 2.87) \times 10^{-3} \, ,
\end{equation}
where indeed the decay of $B^{0}_{s}\rightarrow \phi \pi^{+} \pi^{-}$ is more suppressed than the one of $B^{0}_{s}\rightarrow J/\psi \pi^{+} \pi^{-}$.  And, within our theoretical model, we have
\begin{equation}
\frac{\text{Br}(B^{0}_{s} \rightarrow \phi f_{0}(980))}{\text{Br}(B^{0}_{s}\rightarrow J/\psi f_{0}(980))}=(\frac{V_{p}}{V_{p}^{\prime}})^{2} \times 3.78 \, .
\end{equation}
Therefore, we can obtain $(\frac{V_{p}}{V_{p}^{\prime}})^{2} = (2.31 \pm 0.76) \times 10^{-3}$. Moreover, this value is similar for both cases with or without the contributions of the $\eta\eta$ channel. The predicted ratios using the value of $\frac{V_{p}}{V_{p}^{\prime}}$ are presented in Table \ref{tab:tab3}, where, again, the first uncertainty is relevant to the experimental results and the second one corresponds to the limits of the integration.
Based on these results, using the experimental branching ratio of $\text{Br}(B^{0}\rightarrow J/\psi f_{0}(500))= 8 ^{+1.1}_{-0.9} \times 10^{-6}$, we can get the branching fraction of $B^{0}\rightarrow \phi f_{0}(500)$: $\text{Br}(B^{0}\rightarrow \phi f_{0}(500))= (5.67^{+2.64}_{-2.50}$ $^{+0.03}_{-0.02} )\times 10^{-8}$ (with the $\eta\eta$ channel) and $\text{Br}(B^{0}\rightarrow \phi f_{0}(500))= (5.67^{+2.64}_{-2.50}$ $^{+0.02}_{-0.02} )\times 10^{-8}$  (without the $\eta\eta$ channel), which are consistent the results obtained in Table \ref{tab:tab1} within the uncertainties.

\begin{table}
\renewcommand{\arraystretch}{1.7}
     \setlength{\tabcolsep}{0.2cm}
\center
\caption{Predictions for the branching ratios.}
\resizebox{0.75\textwidth}{!}{\begin{tabular}{|c|c|c|}
\hline
   Ratios & Without $\eta\eta$ channel  & With $\eta\eta$ channel    \\ \hline 
   $\frac{\text{Br}(B^{0}\rightarrow \phi f_{0}(980))}{\text{Br}(B^{0}\rightarrow J/\psi f_{0}(980))}$  &    $(8.35 \pm 2.74  _{-0.16}^{+0.24}) \times 10^{-3}$     & $(8.33 \pm 2.73_{-0.13}^{+0.16}) \times 10^{-3}$   \\ \hline   
   $\frac{\text{Br}(B^{0}\rightarrow \phi f_{0}(500))}{\text{Br}(B^{0}\rightarrow J/\psi f_{0}(500))}$   &   $(7.08 \pm 2.32 _{-0.03}^{+0.03}) \times 10^{-3}$ & $(7.09 \pm 2.32 _{-0.03}^{+0.03}) \times 10^{-3}$   \\ \hline
\end{tabular}}
\label{tab:tab3}
\end{table}

Furthermore, for the $B^{0}_{s} \rightarrow \phi\rho^{0}$ decay, similarly we can determine the value of the vertex factor of $\tilde{V}_{P}$ by the experimental branching fraction of $B^{0}_{s} \rightarrow \phi\rho^{0}$, $\text{Br}(B^{0}_{s} \rightarrow \phi \rho^{0})=(2.7\pm 0.8) \times 10^{-7}$ \cite{pdg2018}. Using Eq. \eqref{eq13},  we have
\begin{equation}
\text{Br}(B^{0}_{s} \rightarrow \phi\rho^{0})= \frac{\Gamma_{B^{0}_{s} \rightarrow \phi\rho^{0}}}{\Gamma_{B_{s}}}= \frac{\tilde{V}_{P}^{2}}{\Gamma_{B_{s}}} \times 1.14 \times 10^{-12}.
\end{equation}
Thus, using the experimental results as input, we can obtain $\frac{\tilde{V}_{P}^{2}}{\Gamma_{B_{s}}}= (2.36 \pm 0.70) \times 10^{5}$.
On the other hand, also with Eq. \eqref{eq13}, one can determine the ratios for the others of $\phi V$ vector decay channels.  For example, the $\phi\phi$ decay channel,  based on the measured branching fraction of $\text{Br}(B^{0}_{s} \rightarrow \phi \phi)= (1.87 \pm 0.15) \times 10^{-5}$, we can obtain the value of the vertex factor as $\frac{(\tilde{V}_{P}^{\prime})^{2}}{\Gamma_{B_{s}}}= (8.05 \pm 0.65) \times 10^{2}$, 
\footnote{Note that, the vertex factor $\tilde{V}_{P}$ for the decay $ B^{0}_{s} \rightarrow \phi\rho^{0}$ is different from the one $\tilde{V}^{\prime}_{P}$ in the $B^{0}_{s} \rightarrow \phi \phi$ decay, because the decay of $ B^{0}_{s} \rightarrow \phi\rho^{0}$ only has the weak interactions in the intermediate processes, whereas, the case of $B^{0}_{s} \rightarrow \phi \phi$ has the strong and the weak interactions in the intermediate procedures.} 
where the uncertainty comes from the experimental value of the branching ratio. Analogous to the others, the results are related to the CKM matrix elements for the intermediate, and one can easy to get the ratios as below, 
\begin{equation}\begin{array}{l}
R_{1}^{th}=\frac{\Gamma_{B^{0} \rightarrow \phi \rho^{0}}}{\Gamma_{B^{0}_{s} \rightarrow \phi \phi}}=\frac{1}{4} \frac{1}{2}\left|\frac{V_{ub}  V_{u d} +V_{c b}  V_{c d} }{V_{ub}  V_{u s} +V_{c b}  V_{c s} }\right|^{2} \frac{m_{B^{0}_{s}}^{2}}{m_{B^{0}}^{2}} \frac{p_{\rho^{0}}}{p_{\phi}}= 6.03 \times 10^{-3},  \\
R_{2}^{th}=\frac{\Gamma_{B^{0} \rightarrow \phi \omega}}{\Gamma_{B^{0}_{s} \rightarrow \phi \phi}}=\frac{1}{4} \frac{1}{2}\left|\frac{V_{ub}  V_{u d} +V_{c b}  V_{c d} }{V_{ub}  V_{u s} +V_{c b}  V_{c s} }\right|^{2} \frac{m_{B^{0}_{s}}^{2}}{m_{B^{0}}^{2}} \frac{p_{\omega}}{p_{\phi}}= 6.03 \times 10^{-3}, \\
R_{3}^{th}=\frac{\Gamma_{B^{0}_{s} \rightarrow \phi \bar{ K}^{* 0}}}{\Gamma_{B^{0}_{s} \rightarrow \phi \phi}}= \left|\frac{V_{ub} V_{u d} +V_{c b}  V_{c d} }{V_{ub}  V_{u s} +V_{c b}  V_{c s} }\right|^{2} \frac{p_{\bar{ K}^{* 0}}}{p_{\phi}}= 4.72 \times 10^{-2}.
\label{eq:ratio1}
\end{array}\end{equation}
The only available experimental ratio \cite{pdg2018} is
\begin{equation}
R_{3}^{exp}= \frac{\text{Br}(B^{0}_{s} \rightarrow \phi \bar{ K}^{* 0})}{\text{Br}(B^{0}_{s} \rightarrow \phi\phi)} = \frac{(1.14 \pm 0.30) \times 10^{-6} }{(1.87 \pm 0.15) \times 10^{-5}}= (6.09 \pm 2.09) \times 10^{-2} \textbf{ },
\end{equation}
where, we can see that our predicted $R_{3}^{th}$ is consistent with the experimental results within the uncertainties.
Besides, using the determined vertex factors above, we can also obtain the other three branching ratios,
\begin{equation}\begin{array}{l}
\text{Br}(B^{0} \rightarrow \phi \rho^{0}) = \frac{\Gamma_{B^{0} \rightarrow \phi\rho^{0}}}{\Gamma_{B}}= (1.13 \pm 0.09) \times 10^{-7},\\
\text{Br}(B^{0} \rightarrow \phi \omega)=\frac{\Gamma_{B^{0} \rightarrow \phi \omega}}{\Gamma_{B}}= (1.13 \pm 0.09) \times 10^{-7}, \\
\text{Br}(B^{0}_{s} \rightarrow \phi \bar{ K}^{* 0})=\frac{\Gamma_{B^{0}_{s} \rightarrow \phi \bar{ K}^{* 0}}}{\Gamma_{B_{s}}}= (8.83 \pm 0.71) \times 10^{-7},
\label{eq:ratio2}
\end{array}\end{equation}
which are consistent with the experimental results \cite{pdg2018} within the upper limits,
\begin{equation}\begin{array}{l}
\text{Br}(B^{0} \rightarrow \phi \rho^{0})< 3.3 \times 10^{-7},\\
\text{Br}(B^{0} \rightarrow \phi \omega)< 7 \times 10^{-7}, \\
\text{Br}(B^{0}_{s} \rightarrow \phi \bar{ K}^{* 0})= (1.14 \pm 0.30) \times 10^{-6}.
\end{array}\end{equation}
As one can see that, for the case of $B^{0}_{s} \rightarrow \phi \bar{ K}^{* 0}$, the predicted value for the branching ratio is in agreement with the experiment within the uncertainties.

\section{Conclusions}

The rare non-leptonic three body decays of $B^{0}_{s} \rightarrow \phi\pi^{+}\pi^{-}$ and $B^{0} \rightarrow \phi\pi^{+}\pi^{-}$, which induced by the flavor changing neutral current $b\rightarrow s\bar{s}s$ and $b\rightarrow d\bar{s}s$, respectively, are studied with the final state interaction approach, based on the chiral unitary approach, where the contributions from the scalar resonances ($f_{0}(500)$ and $f_{0}(980)$) and vector mesons ($\rho$, $\omega$, $\phi$, and $ \bar{ K}^{* 0}$) are taken into account in the final state interactions. Our results for the $\pi^+\pi^-$ invariant mass distributions of the $B^{0}_{s}\rightarrow \phi\pi^{+}\pi^{-}$ decay describe the experimental data up to 1 GeV well when we consider two resonances contributions of the $f_{0}(980)$ and $\rho$, whereas, there is no clear contributions of the $f_{0}(500)$ state in our formalism as indicated in the experiments. Based on these results, we make a prediction for the mass spectrum of the $B^{0}$ decay, where we found that the contributions from the $f_{0}(500)$ state are larger than the one of the $f_{0}(980)$ in the decay of $B^{0} \rightarrow \phi\pi^{+}\pi^{-}$, where an abroad resonance structure can be easily seen in the $\pi^+\pi^-$ invariant mass distributions and a small narrow peak corresponded to the $f_{0}(980)$ also can be found. From these results, one can conclude that the dominant components are the $\pi\pi$ parts in the $f_{0}(500)$ resonance and the $f_{0}(980)$ state is mainly contributed by the $K\bar{K}$ components. Furthermore, we also investigate the branching ratios for the different decay processes with the scalar and vector meson productions in the final states, where some of our results are in agreement with the experiments. Besides, we study the ratios between the $B^0_{(s)}$ decaying into $\phi$ plus the other states and into $J/\psi$ plus the same states. All the predicted results can be seen in Tables \ref{tab:tab1}, \ref{tab:tab3} and Eqs. \eqref{eq:ratio1}, \eqref{eq:ratio2}. Finally, we hope our predicted the $\pi^+\pi^-$ invariant mass distributions for the decay of $B^{0} \rightarrow \phi\pi^{+}\pi^{-}$ and some other branching ratios can be measured by the future experiments.

Note added: When our work is ready, we find that the work of \cite{Zou:2020dpg} also investigate the decays of $B^{0}_{(s)} \rightarrow \phi\pi^{+}\pi^{-}$ with the perturbative QCD approach, which focuses on the branching fractions, the CP asymmetries, and so on.

\section*{Acknowledgments}

We thank E. Oset, J. J. Xie and E. Wang for useful discussions and valuable comments. Z. F. is suported by the National Natural Science Founadtion of China (NSFC) under Grants No. 11705069, and partly suported by NSFC under Grants No. 11965016.

\begin{appendices}

\section{CKM matrix}
\label{ckm}

The CKM matrix elements are fundamental parameters of the SM.The elements of the CKM matrix have been determined from experiments, which can be expressed according to the $A$, $\rho$, $\lambda$, and $\eta$ parameters, called the Wolfenstein parameterization \cite{pdg2018,Wolfenstein:1964ks},
\begin{equation}
V_{\mathrm{CKM}}=\left(\begin{array}{ccc}1-\lambda^{2} / 2 & \lambda & A \lambda^{3}(\rho-i \eta) \\ -\lambda & 1-\lambda^{2} / 2 & A \lambda^{2} \\ A \lambda^{3}(1-\rho-i \eta) & -A \lambda^{2} & 1\end{array}\right)+\mathcal{O}\left(\lambda^{4}\right),
\label{eq:ckm1}
\end{equation}
where the values of these parameters are given by \cite{pdg2018}
\begin{equation}
\begin{array}{l}\lambda=0.22453 \pm 0.00044, \quad A=0.836 \pm 0.015, \quad \bar{\rho}=0.122_{-0.017}^{+0.018}, \quad \bar{\eta}=0.355_{-0.011}^{+0.012}\end{array}, 
\end{equation}
having
\begin{equation}
\bar{\rho}=\rho\left(1-\frac{\lambda^{2}}{2}\right), \quad \bar{\eta}=\eta\left(1-\frac{\lambda^{2}}{2}\right)
\end{equation}
Besides, the absolute value of the CKM matrix including the uncertainty can be given by
\begin{equation}
|V_{\mathrm{CKM}}|=\left(\begin{array}{ccc}0.97446 \pm 0.00010 & 0.22452 \pm 0.00044 & 0.00365 \pm 0.00012 \\ 0.22438 \pm 0.00044 & 0.97359+0.00010 & 0.04214 \pm 0.00076 \\ 0.00896_{-0.00023}^{+0.00024} & 0.04133 \pm 0.00074 & 0.999105 \pm 0.000032 \\ \end{array}\right).
\label{eq:ckm2}
\end{equation}

\section{Formalism of the B meson to $J/\psi \pi^{+}\pi^{-}$}
\label{section:app1}

Following the work of Ref. \cite{Liang:2014tia}, the details for the study of $B\to J/\psi \pi^{+}\pi^{-}$ are summarized as follow,
\begin{equation}
\begin{aligned} 
B^{0}(\bar{b}d) &\Rightarrow [V_{cb}] \bar{c} \mathit{W^{+}} d \Rightarrow [V_{cb}][V_{cd}^{*} ]\textbf{ } (c\bar{c})\, (d\bar{d}) \\
&\Rightarrow [V_{cb}][V_{cd}^{*} ]\textbf{ } (c\bar{c}\to J/\psi)  \textbf{ } [d\bar{d}\to d\bar{d}\cdot (u\bar{u}+d\bar{d}+s\bar{s})]  \\
& \Rightarrow [V_{cb}][V_{cd}^{*} ]\textbf{ } (c\bar{c}\to J/\psi)  \textbf{ } [M_{22}\to (M\cdot M)_{22}] \, ,
\end{aligned} 
\end{equation}
\begin{equation}
\begin{aligned} 
B^{0}_{s} (\bar{b}s) &\Rightarrow [V_{cb}] \bar{c} \mathit{W^{+}} s\Rightarrow [V_{cb}][V_{cs}^{*} ]\textbf{ } (c\bar{c})\, (s\bar{s}) \\
&\Rightarrow [V_{cb}][V_{cs}^{*} ]\textbf{ } (c\bar{c}\to J/\psi)  \textbf{ } [s\bar{s} \to s\bar{s}\cdot (u\bar{u}+d\bar{d}+s\bar{s})] \\
&\Rightarrow [V_{cb}][V_{cs}^{*} ]\textbf{ } (c\bar{c}\to J/\psi)  \textbf{ } [M_{33}\to (M\cdot M)_{33}] \, ,
\end{aligned} 
\end{equation}
where the matrix $M$ is defined in Eq. \eqref{eq:matrM}. Thus, for the hadronization procedures we have
\begin{equation}
\begin{aligned}  
d \bar{d} \cdot (u \bar{u}+d \bar{d}+s \bar{s}) & \equiv(\Phi \cdot \Phi)_{22} =\pi^{+} \pi^{-}+\frac{1}{2} \pi^{0} \pi^{0}-\frac{1}{\sqrt{3}} \pi^{0} \eta+K^{0} \bar{K}^{0}+\frac{1}{6} \eta \eta ,\\ 
s \bar{s} \cdot (u \bar{u}+d \bar{d}+s \bar{s}) & \equiv(\Phi \cdot \Phi)_{33}=K^{-} K^{+}+K^{0} \bar{K}^{0}+\frac{4}{6} \eta \eta,  
\end{aligned}
\end{equation} 
with the matrix $\Phi$ given in Eq. \eqref{eq:Mphi}.

The amplitudes for $\pi^{+}\pi^{-}$ productions are given by
\begin{equation}
\begin{aligned} 
t\left(B^{0} \rightarrow J/\psi \pi^{+} \pi^{-}\right) =& V^{\prime}_{P} (V_{cb} V_{cd}^{*} )\left(1+G_{\pi^{+} \pi^{-}} t_{\pi^{+} \pi^{-} \rightarrow \pi^{+} \pi^{-}} \right. +2 \frac{1}{2} \frac{1}{2} G_{\pi^{0} \pi^{0}} t_{\pi^{0} \pi^{0} \rightarrow \pi^{+} \pi^{-}} \\
&+G_{K^{0} \bar{K}^{0}} t_{K^{0} \bar{K}^{0} \rightarrow \pi^{+} \pi^{-}}\left.+2\frac{1}{6} \frac{1}{2} G_{\eta \eta} t_{\eta \eta \rightarrow \pi^{+} \pi^{-}}\right), \\ 
t\left(B^{0}_{s} \rightarrow J/\psi \pi^{+} \pi^{-}\right) =& V^{\prime}_{P} (V_{cb} V_{cs}^{*} )\left( G_{K^{+}K^{-}}t_{K^{+}K^{-} \rightarrow \pi^{+} \pi^{-}}\right. +G_{K^{0} \bar{K}^{0}} t_{K^{0} \bar{K}^{0} \rightarrow \pi^{+} \pi^{-}} \\
& \left.+2\frac{4}{6} \frac{1}{2} G_{\eta \eta} t_{\eta \eta \rightarrow \pi^{+} \pi^{-}}\right), 
\end{aligned}
\end{equation}
Finally, the partial decay widths can be written as
\begin{equation}
\frac{d \Gamma}{d M_{\text{inv}}}=\frac{1}{(2 \pi)^{3}} \frac{1}{8 M_{B_{(s)}}^{2}} \frac{2}{3} p_{J/\psi}^{2} p_{J/\psi} \tilde{p}_{\pi} \bar{\sum} \sum \left|t_{B_{(s)}^{0} \rightarrow J/\psi \pi^{+} \pi^{-}}\right|^{2}.
\end{equation} 
Thus, when the $\eta \eta$ channel is considered in the coupled channel interactions, we have
\begin{equation} 
\text{Br}(B^{0}_{s} \rightarrow J/\psi f_{0}(980))= \frac{\Gamma_{B^{0}_{s} \rightarrow J/\psi f_{0}(980)}}{\Gamma_{B_{s}}}= \frac{\int_{2m_{\pi}}^{1200}\frac{d\Gamma_{B^{0}_{s} \rightarrow J/\psi f_{0}(980)}}{dM_{inv}}dM_{inv}}{\Gamma_{B_{s}}}  \\ = \frac{V_{P}^{\prime 2}}{\Gamma_{B_{s}}} \times 8.66 \textbf{ },
\label{eq:49}
\end{equation} 
and ignored the $\eta \eta$ channel in the two-body interactions,
\begin{equation} 
\text{Br}(B^{0}_{s} \rightarrow J/\psi f_{0}(980))= \frac{\Gamma_{B^{0}_{s} \rightarrow J/\psi f_{0}(980)}}{\Gamma_{B_{s}}}= \frac{\int_{2m_{\pi}}^{1200}\frac{d\Gamma_{B^{0}_{s} \rightarrow J/\psi f_{0}(980)}}{dM_{inv}}dM_{inv}}{\Gamma_{B_{s}}}  \\ = \frac{V_{P}^{\prime 2}}{\Gamma_{B_{s}}} \times 13.54 \textbf{ }.
\label{eq:50}
\end{equation}
Using the measured branching fraction of the $\text{Br}(B^{0}_{s} \rightarrow J/\psi f_{0}(980))= (1.28 \pm 0.18) \times 10^{-4}$ \cite{pdg2018}, we can obtain $ \frac{V_{P}^{\prime 2}}{\Gamma_{B_{s}}}= (1.48 \pm 0.21) \times 10^{-5}$ and $\frac{V_{P}^{\prime 2}}{\Gamma_{B_{s}}}= (9.46 \pm 1.33) \times 10^{-6}$ for with and without the $\eta \eta$ channel in the two-body interactions, respectively.

\begin{figure}
  \centering
  \includegraphics[width=0.6\linewidth]{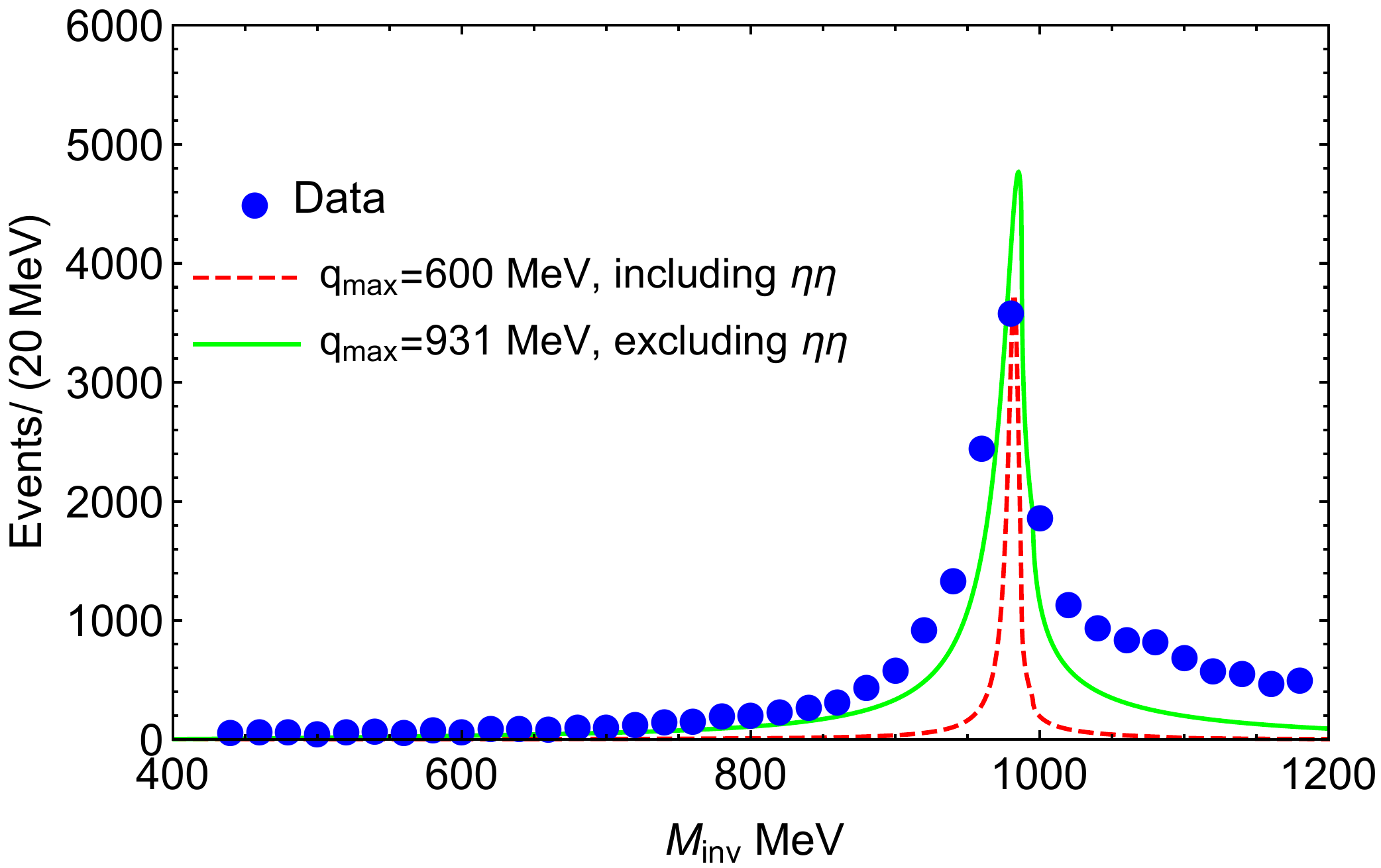}
\caption{$\pi^{+}\pi^{-}$ invariant mass distributions of the $B^{0}_{s}\rightarrow J/\psi\pi^{+}\pi^{-}$ decay, where we plot $\frac{C^\prime \times 10^{-8}}{\Gamma_{B_{s}}}  \frac{d \Gamma}{d M_{\text{inv}}}$ and the data is taken from Ref. \cite{Aaij:2014emv}. The dash (red) and solid (green) lines correspond to the results with (the normalization constant $C^\prime=5.36$) and without ($C^\prime=19.71$) the $\eta \eta$ channel in the two-body interactions, respectively.}
\label{fig10}
\end{figure}

\begin{figure}
  \centering
  \includegraphics[width=0.6\linewidth]{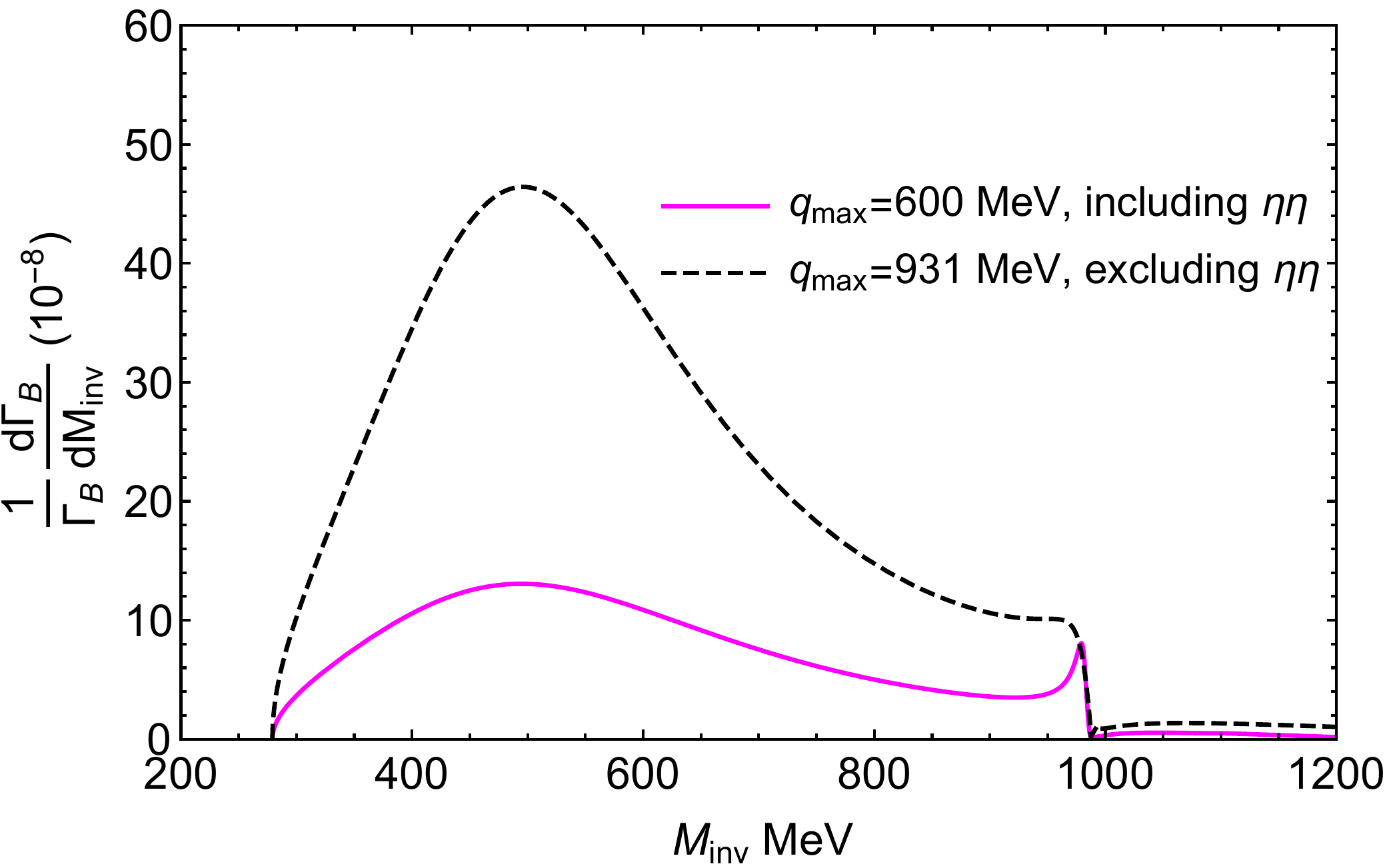}
\caption{$\pi^{+}\pi^{-}$ invariant mass distributions for the $B^{0}\rightarrow J/\psi\pi^{+}\pi^{-}$ decay, where the solid (magenta) and dash (black) lines represent the results with and without the coupled channel of $\eta \eta$, respectively. }
\label{fig11}
\end{figure}

The $\pi^{+}\pi^{-}$ invariant mass distributions of $B^{0}_{s}\rightarrow J/\psi \pi^{+} \pi^{-}$ and $B^{0}\rightarrow J/\psi \pi^{+} \pi^{-}$ are shown in Fig. \ref{fig10} and Fig. \ref{fig11}, respectively, which are consistent with the ones of Ref. \cite{Liang:2014tia}.

\end{appendices}

 \addcontentsline{toc}{section}{References}
\end{document}